\documentclass{elsarticle}

\usepackage{graphicx}      
\usepackage{amsfonts}
\usepackage{amsmath}
\usepackage{amssymb}
\usepackage{makeidx}
\usepackage{epsfig}
\usepackage{graphicx}
\usepackage{epstopdf}

\usepackage{natbib}

\newdefinition{rmk}{Remark}
\newproof{pf}{Proof}
\newproof{pot}{Proof of Theorem \ref{thm2}}
\newdefinition{assumption}{Assumption}
\newdefinition{problem}{Problem}
\newdefinition{definition}{Definition}
\newdefinition{proposition}{Proposition}

\begin{document}
\baselineskip 10.5pt
%
%
%
\begin{frontmatter}
\title{\LARGE Safe Human-Inspired Mesoscopic Hybrid Automaton for Autonomous Vehicles}
%
%

%
\author[DISIMDEWS]{A. Iovine\corref{cor1}\fnref{fn1}}
\cortext[cor1]{Corresponding author}
\ead{alessio.iovine@graduate.univaq.it}
\fntext[fn1]{Permanent email: alessio.iovine@hotmail.com}
\author[DISIMDEWS]{F. Valentini}
\ead{francesco.valentini1@graduate.univaq.it}
\author[DISIMDEWS]{E. De Santis}
\ead{elena.desantis@univaq.it}
\author[DISIMDEWS]{M. D. Di Benedetto}
\ead{mariadomenica.dibenedetto@univaq.it}
\author[DISIMDEWS]{M. Pratesi}
\ead{marco.pratesi@univaq.it}

\address[DISIMDEWS]{Department of Information Engineering, Computer Science and Mathematics \\DEWS Center of Excellence\\
University of L'Aquila, Via Vetoio, Coppito, 67100, Italy.}
%
%
%
\begin{abstract}                
In this paper a mesoscopic hybrid model, i.e. a microscopic hybrid model that takes into account macroscopic parameters, is introduced for designing a human-inspired Adaptive Cruise Control. A control law is proposed with the design goal of replacing and imitating the behaviour of a human driver in a car-following situation where lane changes are possible. First, a microscopic hybrid automaton model is presented, based on human psycho-physical behavior, for both longitudinal and lateral vehicle control. Then a rule for changing time headway on the basis of macroscopic quantities is used to describe the interaction among next vehicles and their impact on driver performance. Simulation results show the advantages of the mesoscopic model. A feasibility analysis of the needed communication network is also presented.
\end{abstract}%
\end{frontmatter}%
\section{Introduction}\label{Introduction}%
Traffic control is one of the most studied problems in engineering because of its impact on human life: progress in the knowledge and control of traffic systems would improve life quality (see \cite{hoogendoorn2001state}).
The goal of traffic control is to manage the flow of cars in highways so that a number of quantities such as congestion, emissions, travel time reduction and safety are traded-off in an optimal fashion.  Driver support systems such as Adaptive Cruise Control (ACC) systems, and Advanced Driver-Assistance Systems (ADAS) provide full or partial driver assistance (\cite{Varaiya1993SmartCars}, \cite{Ioannou1993ACC}, \cite{Bajcsy2014Semiautonomous}) to control traffic. Once ADAS systems are introduced in cars, the behavior of the cars should follow normal traffic dynamics. To do so, we believe driver support systems should mimic driver behaviour. 

The objective of this paper is developing an ACC model able to mimic the human driver behavior with respect to comfort while ensuring a proper safety level. With this intent, classical models will be used. Over the years, a multitude of traffic control systems have been proposed (see \cite{hoogendoorn2001state}, \cite{Panwai2005-TITS}). Those models can be classified on the basis of their level of abstraction (macroscopic, as in \cite{Kotsialos1999ACC}, \cite{Papageorgiou2002metanet}, microscopic as in \cite{Gazis1961FollowTheLeader}, \cite{GIPPS1981Behavioural} or mesoscopic, a microscopic model that takes into account macroscopic parameters) or the adopted control strategy (centralized, as in \cite{Daganzo}, or decentralized as in \cite{falconi2012hybrid}, \cite{Caravani2006TITS}) and have been utilized with different methodologies for managing a large variety of problems (\cite{Heemels2012StringStability}, \cite{Papageorgiou1990OptimalControl}, \cite{Xiao2011StringStability}, \cite{DiBenedetto2013Grenoble}, \cite{NaiOleari2015MPCregressionTrees}, \cite{Tabuada2015ACC}). 

In this paper, we present a decentralized mesoscopic control approach for developing an ACC system. In \cite{Iovine2015ADHSofficial}, we proposed a hybrid automaton model for regulating vehicle interactions on a single lane road. In this paper, we extend our modeling framework to the automatic control of a vehicle in a multilane road with overtake possibility.
We include new situations with the goal of imitating a human driver while avoiding model complexity explosion. To take into consideration the needed information, the proposed hybrid automaton includes additional states and external events, which are related to the new control actions, their robustness and safety requirements. Safety and stability properties are proven when controlling the entire group of vehicles on the multilane road, whereby the single lane case developed in \cite{Iovine2015ADHSofficial} becomes a special case.

We consider $N$ vehicles, indexed by $n\in\left\{1,...,N\right\}$, on a road with two lanes.
Vehicle $n$ can be in a lane-maintaining mode or in a lane-change mode. Consider the closest vehicle ahead in the same lane as $n$ and, if any, the one in change-lane mode that is closest to $n$ in the other lane. The "leader" for $n$ is the vehicle closest to $n$ between the two. Once the leader has been identified, the vehicle $n$ will consider itself a "follower".
A hybrid model describing a "leader" and a "follower" pair is developed, based on classical psycho-physical and stimulus-response car-following models. We will show that this  hybrid automaton is deterministic and satisfies non-blocking and safety properties.

\begin{figure}
	\centering\includegraphics[width=1\columnwidth]{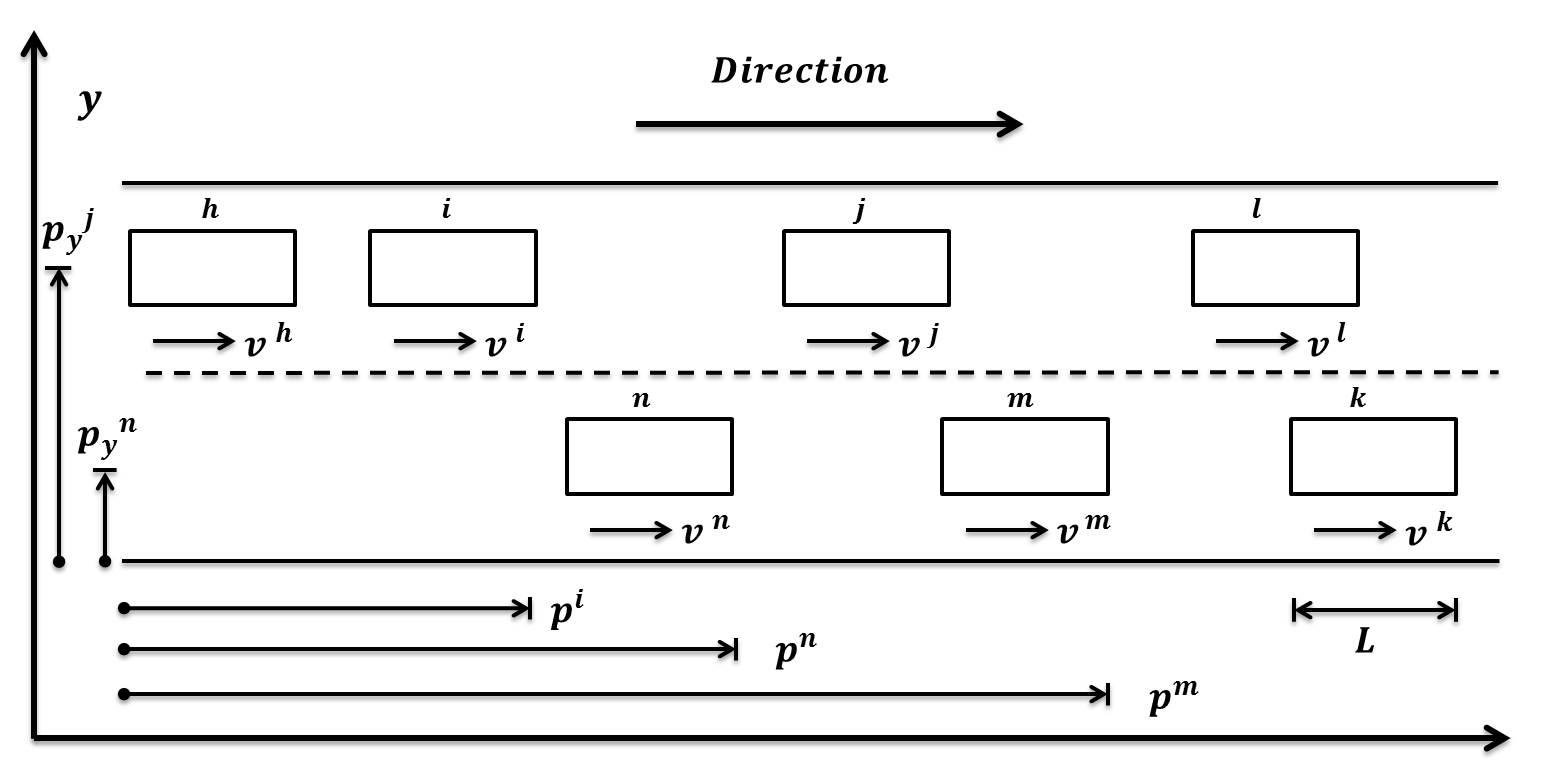}
	\caption{The reference framework of a road with two lanes, the $r$ right one and the $l$ left one.}\label{Figure_reference}
\end{figure}%
%
We first analyze the properties of the overall hybrid system when applying state feedback control laws, using only local information. 
Such control laws simulate the human control action (following the models of human behavior illustrated in the car-following models), where the objective is to minimize traveling time while maintaining safety.
In the second part of the paper, we define a mesoscopic model, where the control action depends not only on individual information but also on the traffic flow, which is a macroscopic quantity. The macroscopic information can either be provided by a centralized traffic supervisor, or it can be gathered, elaborated and transmitted by the vehicles themselves, when they are connected and can exchange information (see \cite{Uhlemann}). We consider the second case.
We will show how this "decentralized" architecture reduces acceleration peaks and improves the throughput of the highway system. Further we will show how the use of mesoscopic information influences lane change. State minimization is a major goal in developing the model to prevent complexity explosion. Some simulation results are offered, as well as a discussion on communication questions regarding the proposed model feasibility.

This paper is organized as follows. 
In Section \ref{Microscopic Hybrid Model} the model of the microscopic hybrid automaton for a single vehicle is described. 
Then in Section \ref{Mesoscopic model} a variance-driven time headway
mechanism is introduced in the hybrid automaton, which thereby becomes
mesoscopic. 
Then Section \ref{vehicular_network} presents the communication framework, while Section \ref{Simulation results} provides simulation results about the system behaviour, which demonstrate the advantages of the proposed mesoscopic model. Conclusions are offered in Section \ref{Conclusions}.

\section{Problem definition}\label{Problem_definition}
We consider a cluster of $N$ vehicles, travelling in the same direction along a road with two lanes.

\begin{figure}
	\centering\includegraphics[width=0.9\columnwidth]{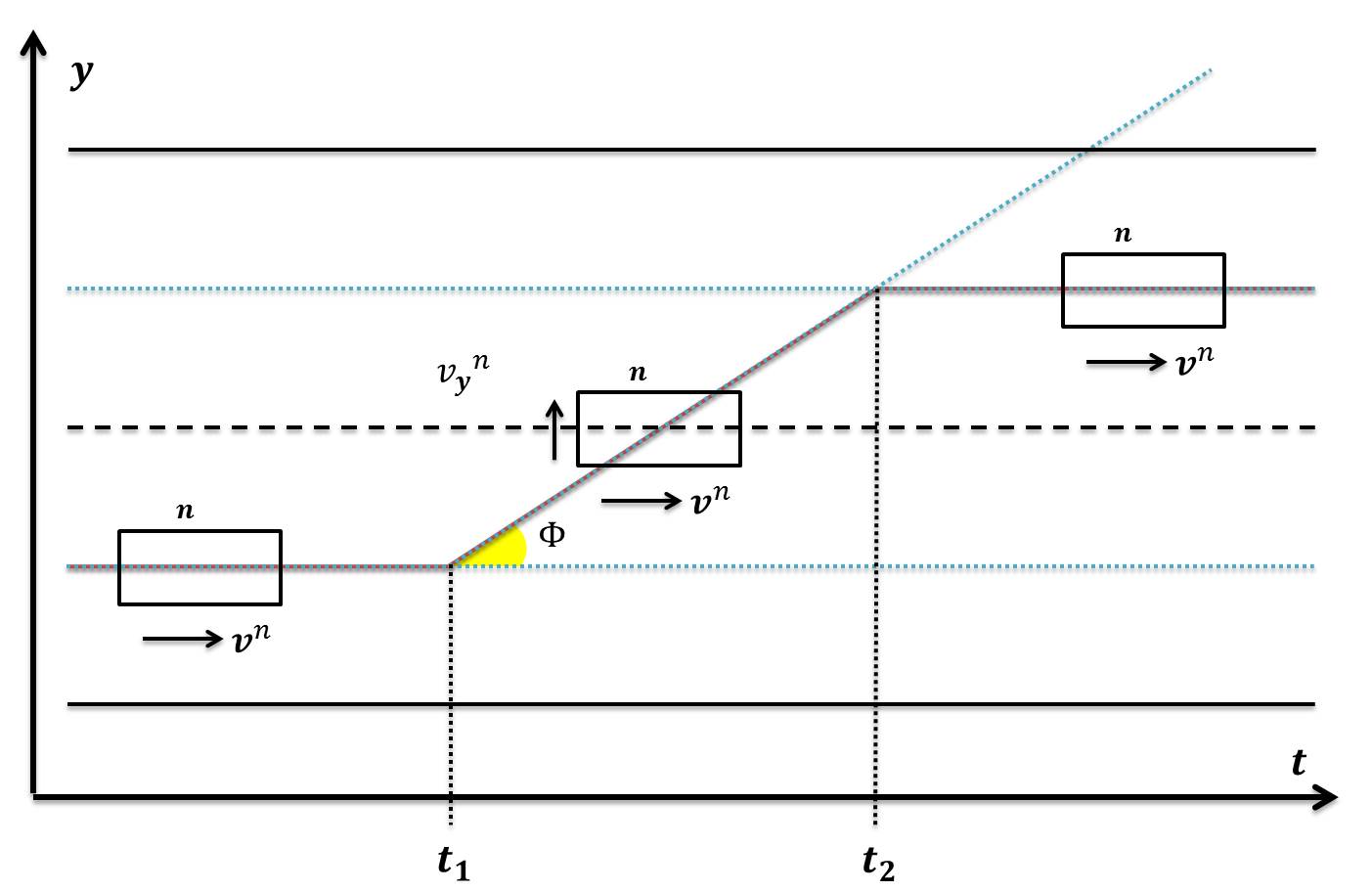}
	\caption{The lane change over time. The monodirectional motion of vehicle $n$ becomes bidirectional at $t=t_1$; at $t=t_2$ the desired point of the target lane is reached and there is a switching to again a monodirectional motion.}\label{Figure_lane_change}
\end{figure}

Both lane maintain and lane change will be considered. As depicted in Figure \ref{Figure_reference}, when in lane maintain mode, the physical dynamics is described on the longitudinal axis only. For physically actuating the lane change the dynamics is defined on two axes. When the desired value on the $\vec{y}$ axis is reached, the motion is again only developed on the longitudinal axis. For simplicity, we suppose that switching among the two dynamics is instantaneous and actuated with a $\phi$ angle (see Figure \ref{Figure_lane_change}).

We assume that all the vehicles have the same length $L$. Given the vehicle $n$, $n=1,...,N$, its position and velocity  on the longitudinal (resp. transversal) axis are denoted by ${p^{n}(t)}$ and ${v^{n}(t)}$ (resp. ${p_{\mathbf{y}}^{n}(t)}$ and ${v_{\mathbf{y}}^{n}(t)}$), ${p^{n}(t)\geq0}$, ${p_{\mathbf{y}}^{n}(t)}\geq0$, $0\leq{v^{n}(t)}\leq v_{max}$, $0\leq{v_{\mathbf{y}}^{n}(t)}\leq v_{max}$. Let us define what a collision is.

\begin{definition}\label{collision_def}
Given two vehicles $n$ and $m$ in the same lane, a \emph{collision} between $n$ and $m$ at time $t$ is defined as the event corresponding to a longitudinal distance between $n$ and $m$ less than $s$, which is the sum of the vehicle length $L$ and a minimum distance $L_{0}$, i.e. $\left\vert p^{n}(t)-p^{m}(t)\right\vert \leq s=L+L_{0}$.
\end{definition}
We make the following assumption about rationality of the cluster:
\begin{assumption}\label{self}
All vehicles in the cluster have the same goal of avoiding collisions.
\end{assumption}

\begin{definition}
The situation where the deceleration of one or more vehicles in the cluster is maximum is called worst case scenario.
\end{definition}

Suppose that each vehicle $n$ has a desired constant velocity $v_{des}^{n}$. Then the problem we address in this paper is:

\begin{problem} \label{problem}
Design decentralized control laws such that collisions are avoided in the worst case scenario and the velocity of vehicle $n$ is as close as possible to $v_{des}^{n}$, $\forall n=1,...,N$.
\end{problem}

As we will better describe in Section \ref{vehicular_network}, we assume that information can be exchanged among the vehicles, by using WAVE (Wireless Access in Vehicular Environment) protocol stack. Therefore we denote with
$\Delta_{max}$ the Radio Range, i.e. the maximum distance such that communication between two vehicles is possible with a sufficiently high quality.

Problem \ref{problem} will be solved by considering two different information frameworks.
In the first case, each vehicle has information about its neighbors as follows:

\begin{definition}\label{def_neighbor}
Given $n$, let $\mathcal{R}_n(t)$ be the set of vehicles whose distance from $n$ at time $t$ is less than or equal to $\Delta_{max}$. The neighborhood of $n$ at time $t$, denoted by $\mathcal{I}_n(t)$, is the set of vehicles $\left\{m,i,j\right\}\subset  \mathcal{R}_n(t)$, where $m$ is the closest vehicle ahead in the same lane, $i$ is the closest behind in the other lane (i.e. such that $p^{n}(t)-p^{i}(t)>0$) and $j$ is the closest ahead in the other lane (i.e. such that $p^{j}(t)-p^{n}(t)\geq0$).
\end{definition}

The neighborhood of $n$ can be empty. Moreover, if the vehicle $i$ is not defined (i.e. there is no vehicle behind in the set $\mathcal{R}_n(t)$), the neighborhood of $n$ will be $\left\{m,\epsilon,j\right\}$ (similarly for the other vehicles $m$ and $j$).

The model based on this local information will be called "microscopic".

In the second case the vehicle has information about its neighbors and also about all the vehicles ahead, up to a given distance. The model based on this second case will be called "mesoscopic".

The improvements due to the use of more information in a mesoscopic framework results in a smoother transient; less oscillations and less magnitude of oscillations provide a better response to problems such as string instability and shock waves propagation.

For each vehicle $n$, we will define a hybrid model whose discrete states are based on the relative speed-distance between $n$ ("the follower") and its leader, defined in the following

\begin{definition}\label{leader_def }
Given the vehicle $n$, let us consider at time $t$ its neighborhood  $\mathcal{I}_n(t)=\{m,i,j\}$. Suppose that $m\neq \epsilon$. If $j=\epsilon$ or if $j$ is not in lane
changing mode, then the leader of $n$ at $t$ is $m$. Otherwise it is the closest vehicle between $m$ and $j$.
\end{definition}

An example of $\mathcal{I}_n(t)=\{m,i,j\}$ is given in Figure \ref{Figure_reference}.

\section{Microscopic Hybrid Model}\label{Microscopic Hybrid Model}
\subsection{Hybrid Automaton}
The problem of modeling the human way of driving has been extensively studied in the literature. We have considered a number of control laws, one for each different situation, and we have embedded them in a unique model, which also takes into account discontinuities due e.g. to changes of lanes or to changes of the neighborhood of the vehicle. A hybrid model is therefore obtained.

We assume that all vehicles are identical. Each vehicle is modeled by a hybrid automaton, affected by a continuous disturbance that represents the control action of the leader and by discrete signals which depend on the behaviour of the vehicles in the cluster. The continuous control law depends on the discrete state, and for a fixed discrete state it is a continuous state feedback. In what follows we describe the hybrid automaton that takes into account the closed loop continuous dynamics. The different continuous control laws will be described later in the section.

The hybrid automaton associated with vehicle $n$,  $n=1,...,N$, is described by the tuple%
\begin{equation}\label{HybridAutomaton}%
\mathcal{H}^{n}=(Q,X,V,f,Init,Dom,\mathcal{E},\mathcal{R})
\end{equation}
where
\begin{itemize}
\item $Q=\left\{q_{i},i=1,...,36\right\}$ is the set of discrete states (also called \emph{modes});
\item $X=\mathbb{R}^{6}$  is the continuous state space;
\item $V=\{\sigma_{nl},\sigma_{ex},\sigma_{c} \}$ is the set of external events;
\item $f=\left\{f_{q},q\in Q\right\}$, and $f_{q}:X\times\mathbb{R}\rightarrow X $ 
    is a  vector field that associates to the discrete state $q\in Q$ the continuous time-invariant dynamics
\begin{equation}
\dot{x}(t)=f_{q}(x(t),d(t))\label{diffeq}%
\end{equation}
where $d:\mathbb{R}\rightarrow\mathbb{R}$  is the control input of the leader vehicle, modeled as a bounded disturbance from the point of view of vehicle $n$, i.e. $\left\vert d(t)\right\vert \leq a_{\max}$;
\item $Init\subseteq Q\times X$ \ is the set of initial discrete and continuous conditions;
\item $Dom:Q\rightarrow2^{X}$; %
\item  $\mathcal{E}\subseteq Q\times  2^{V} \times Q$ \ is the set of edges;
\item  $\mathcal{R}: \mathcal{E}\times X\times\mathbb{R}^2\rightarrow X$ is the reset function.
\end{itemize}
The automaton hybrid state is the pair $(x,q)\in X\times Q$.
\subsection{Continuous states}
%
%
If the leader of vehicle $n$ at time $t$ is well defined, then $\widehat{p}(t)$ and $\widehat{v}(t)$ denote the position and speed of the leader. Otherwise they represent the position and speed of a fictitious vehicle, travelling with speed $\widehat{v}(t)=v_{des}^n$ (i.e. $d(t)=0$), which is "very far" from $n$, i.e. the distance is higher than $\Delta_{max}$. This last condition will be ensured by appropriately initializing the system, as described in Section \ref{subsec_reset}.  Let $\varphi(t)\in \{0,\Phi\}$ be the steering angle (see e.g. Figure \ref{Figure_lane_change} where $\varphi(t)=0$ for $t<t_1$ and $t\geq t_2$, while $\varphi(t)=\Phi$ for $t\in[t_1,t_2)$). Then, the continuous state of ${n}$ is

\begin{equation}\label{eq_continuous_state}
x(t)=\left[
\begin{array}
[c]{c}%
x_{1}(t)\\
x_{2}(t)\\
x_{3}(t)\\
x_{4}(t)\\
x_{5}(t)\\
x_{6}(t)
\end{array}
\right]  =\left[
\begin{array}
[c]{c}%
\widehat{p}(t)-p^{n}(t)\\
\widehat{v}(t)-v^{n}(t)\\
\widehat{v}(t)\\
{p_{\mathbf{y}}^{n}(t)}\\
{v_{\mathbf{y}}^{n}(t)}\\
{\varphi^{n}(t)}
\end{array}
\right]
\end{equation}
\subsection{Discrete states}
At high level, the evolution of the vehicle can be described by
the cycle in (\ref{ciclo}), where label $r$ (resp. $l$) means that the vehicle
is going straight in the right lane (resp. left lane), $r2l$ (resp. $l2r$) means
that the vehicle is in the right lane (resp. left lane) but it is moving to the
left lane (resp. right lane), $r2r$ (resp. $l2l$) means that it is in the right
lane (resp. left lane) coming from the left lane (resp. right lane).
\begin{equation}%
\begin{array}
[c]{ccc}%
l2l & \leftarrow & r2l\\
\downarrow &  & \uparrow\\
l &  & r\\
\downarrow &  & \uparrow\\
l2r & \rightarrow & r2r
\end{array}
\label{ciclo}%
\end{equation}
Let  $Q_{1}=\left\{  r,r2l,l2l,l,l2r,r2r\right\}  $. For each of the high level modes in $Q_{1}$, six different lower level modes are
defined, which only depend on the interaction between vehicle $n$ and its leader. Let $Q_{2}=\left\{w_i, i=1...6\right\}  $  be the set of lower level modes. As we will describe, the continuous state space is partitioned
into $6$ regions $\Gamma_{i}$ and for a given mode $h\in Q_{1}$, the discrete
state is equal to $\left(  h,w_{i}\right)$ at time $t$ whenever
$x(t)\in\Gamma_{i}$. Then
\begin{equation}\label{}
Q=Q_{1}\times Q_{2}%
\end{equation}
\subsection{Domains}
We define the functions ${T_{E}}:X\rightarrow\mathbb{R}$, ${T_{R}}:X\rightarrow\mathbb{R}$ and ${T_{S}}:X\rightarrow\mathbb{R}$, where
\begin{equation}\label{headways}
{T_{E}}({x})=\frac{\left\vert {x_{2}}\right\vert }{a_{\max}}, \:\:\\
{T_{R}}({x})=\frac{\left\vert {x_{3}-x_{2}}\right\vert }{a_{\max}}, \:\:\\
{T_{S}}({x})=\lambda\frac{\left\vert x_{3}-x_{2}\right\vert }{a_{\max}}
\end{equation}
representing, respectively, the time headways needed to stop the vehicle
starting from initial speed $x_{2}$ and $\left\vert x_{3}-x_{2}\right\vert $,
with deceleration $u(t)=-a_{\max}$, and the time needed to stop the vehicle
starting from initial speed $\left\vert x_{3}-x_{2}\right\vert $, with
$u(t)=-\frac{a_{\max}}{\lambda}$, $\lambda>1$. Furthermore, we define ${T_{D}}$ as a fixed time: it is the time headway beyond
which the vehicle can ignore the leader (see \cite{Wiedemann1991},
\cite{Fritzsche1994}, \cite{Iovine2015ADHSofficial}).
\begin{figure}
  \centering
  \includegraphics[width=1\columnwidth]{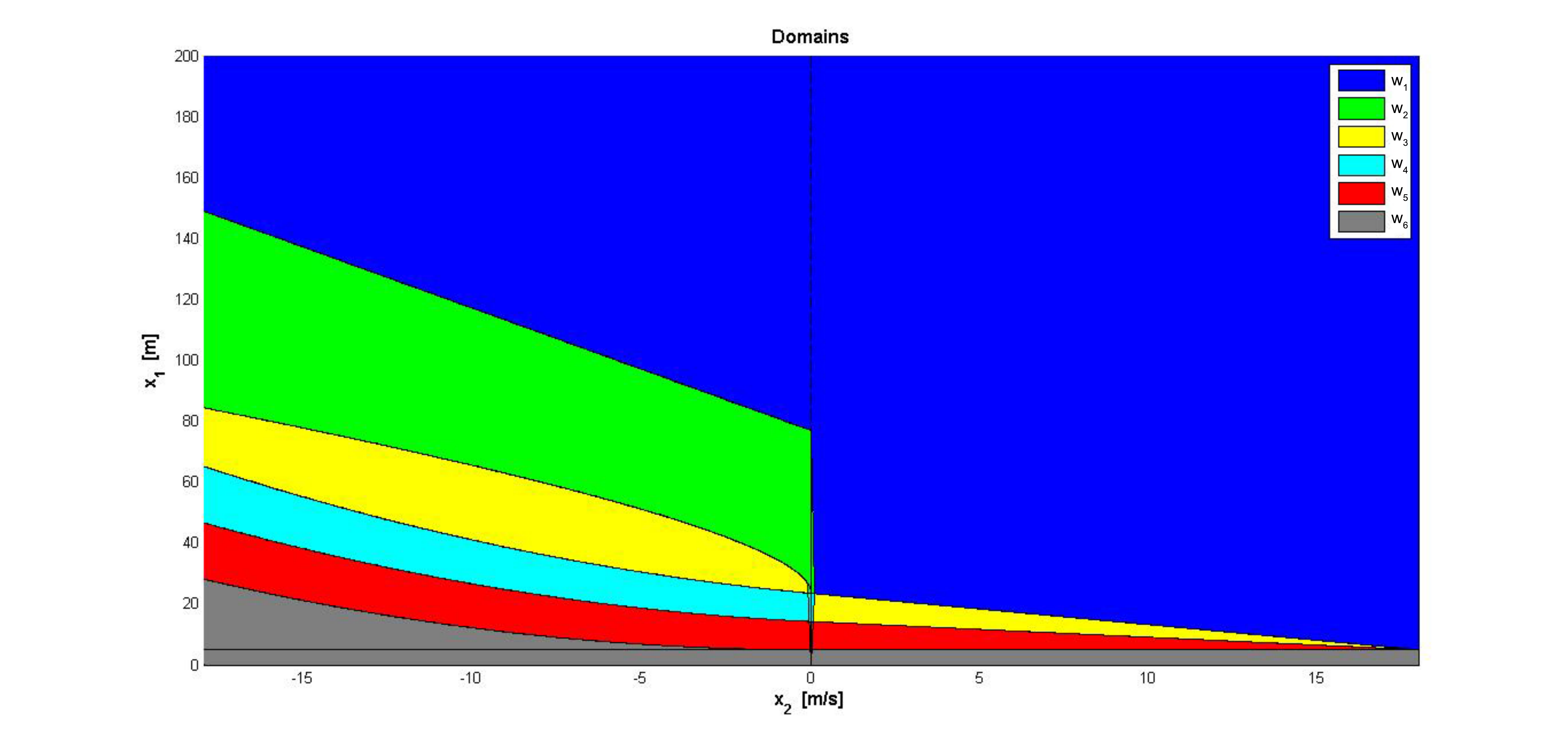}\\
	\caption{The different thresholds and the defined domains at a fixed ${\widehat{v}}=18$ m/s: the $\textit{Free driving}$ domain is in blue, while $\textit{Following I}$ in green, $\textit{Following II}$ in yellow, $\textit{Closing In}$ in cyan, $\textit{Danger} $ in red and the $\textit{Unsafe}$ zone in grey.}\label{Figure_DeltaX_V_diagram}
\end{figure}

We define the following thresholds for
$x_{1}$:
\begin{itemize}
\item \emph{emergency distance} ${\Delta E}:X\rightarrow
\mathbb{R}$ {\footnotesize
\begin{equation}
{\Delta E(x)}=\left\{
\begin{array}
[c]{l}%
s\\
s+\frac{1}{2}a_{\max}{T_{E}}^{2}(x)
\end{array}
\right.
\begin{array}
[c]{c}%
x_{2}>0\\
x_{2}\leq0
\end{array}
\label{Delta_E}%
\end{equation}
}{\normalsize represents the minimum distance such that safety is ensured (see
\cite{GIPPS1981Behavioural}). If at time $t$ the headway is equal to ${\Delta
E(x(t))}$ and the leader starts braking with the maximum deceleration,
provided that the follower also starts braking at the same time with the
maximum deceleration, then collision is avoided. }

\item {\normalsize \emph{risky distance} ${\Delta R}:X%
\rightarrow\mathbb{R}$}{\footnotesize
\begin{equation}
{\Delta R}\left(  x\right)  =\left\{
\begin{array}
[c]{l}%
s+c_{r}{T_{R}}(x)x_{3}\\
s+c_{r}{T_{R}}(x)x_{3}+\frac{1}{2}a_{\max}{T_{E}}^{2}(x)
\end{array}
\right.
\begin{array}
[c]{c}%
x_{2}>0\\
x_{2}\leq0
\end{array}
\label{Delta_R}%
\end{equation}
}{\normalsize has the same interpretation as the distance ${\Delta E\left(
x\right) }$, but it takes into account a human time-response, modeled by the
add-on value $c_{r}{T_{R}}(x)x_{3}$, where $c_{r}>0$ is a constant
multiplication factor.
Depending on the environment information and on the human perception, such
value can increase (more cautious behaviour), or decrease (more aggressive
behaviour), but the condition ${\Delta R\left( x\right) }>{\Delta E(x)}$ is
always satisfied. }

\item {\normalsize \emph{safe distance} ${\Delta S}:X%
\rightarrow\mathbb{R}$}{\footnotesize
\begin{equation}
{\Delta S(x)}=\left\{
\begin{array}
[c]{l}%
s+c_{s}{T_{S}}(x)x_{3}\\
s+c_{s}{T_{S}}(x)x_{3}+\frac{1}{2}a_{\max}{T_{E}}^{2}(x)
\end{array}
\right.
\begin{array}
[c]{c}%
x_{2}>0\\
x_{2}\leq0
\end{array}
\label{Delta_s}%
\end{equation}
}{\normalsize corresponds to an additional safety margin w.r. to ${\Delta R\left(x\right) }$; in fact $c_{s}\geq c_{r}$
and $\lambda>1$ imply that ${T_{S}}>{T_{R}}$ and hence ${\Delta S\left(
x\right) }>{\Delta R\left(  x\right) }$. }

\item {\normalsize \emph{interaction distance} ${\Delta D}:X\rightarrow\mathbb{R}$}{\footnotesize
\begin{equation}\label{Delta_D}%
{\Delta D\left(  x\right) }=\left\{
\begin{array}
[c]{l}%
s+c_{s}{T_{S}}(x)x_{3}\\
s+c_{d}{T_{D}}(x_{3}-x_{2})
\end{array}
\right.
\begin{array}
[c]{c}%
x_{2}>0\\
x_{2}\leq0
\end{array}
\end{equation}
}{\normalsize where ${T_{D}}$ is a fixed time. }

{\normalsize Notice that when $x_{2}>0$, i.e. when ${\widehat{v}(t)}>{v^{n}(t)}$,
${\Delta D(x)}={\Delta S(x)}$. }

\item {\normalsize \emph{approaching distance} ${\Delta C}:X\rightarrow\mathbb{R}$}{\footnotesize
\begin{equation}
{\Delta C(x)}=\left\{
\begin{array}
[c]{l}%
s + c_{s}{T_{S}}(x)x_{3}\\
s + c_{s}{T_{S}}(x)x_{3} + c_{c} \mid\sqrt{-x_{2}}\mid
\end{array}
\right.
\begin{array}
[c]{c}%
x_{2}>0\\
x_{2}\leq0
\end{array}
\label{Delta_C}%
\end{equation}
}{\normalsize is a threshold that corresponds to the driver becoming aware that he is approaching the ahead vehicle with high speed differences at short decreasing distances (see \cite{Wiedemann1991}). }
\end{itemize}

Setting $d(t)=0$, i.e. $\widehat{v}{(t)}$ to a constant ${\mathbf{v}}$, in Figure \ref{Figure_DeltaX_V_diagram} the thresholds are represented in a bidimensional space, where the speed difference ${\mathbf{v}}-{v^{n}}(t)=x_{2}(t)$ is represented on the horizontal axis, while the distance $\widehat{{p}}(t)-{p^{n}}(t)=x_{1}(t)$ is represented on the vertical axis.

The introduced thresholds define the domains of the continuous state space to be associated to the discrete states; here $h\in\{l,r,l2r,l2l,r2l,r2r\}$.
\begin{enumerate}%
\item \textit{Free driving} domain: the vehicle can run freely because the leader vehicle is either too far away or faster or both. $Dom((h,w_{1}))$ is the set
\begin{align}\label{Dom_q1}
\left\{x\in X:(x_{1}>{\Delta S(x)})\wedge(x_{2}\geq0)\right\} \cup\left\{x\in X:(x_{1}>\mathbf{m})\wedge(x_{2}<0)\right\}%
\end{align}
where $\mathbf{m}=\max\left\{  {\Delta D(x)},{\Delta S(x)}\right\}$.

%
\item \textit{Following I} domain: the following vehicle is closing in on the leader vehicle. $Dom((h,w_{2}))$ is the set%
\begin{equation}\label{Dom_q2}
\left\{x\in X:(x_{2}<0)\wedge\\
(\mathbf{n} <x_{1}\leq{\Delta D(x)})\right\}
\end{equation}
with $\mathbf{n}=\max\left\{{\Delta S(x)},{\Delta C(x)}\right\}$.

\item \textit{Following II} domain: the follower does not take any action, either because its speed is close to the leader's
one and the distance is small or because the speed difference is too large with respect to the distance. $Dom((h,w_{3}))$ is the set
\begin{align}\label{Dom_q3}
&\left\{x\in X:(x_{2}\leq0)\wedge({\Delta S(x)}< x_{1}<\mathbf{p})\right\}\cup \\
&\nonumber \left\{ x\in X:(x_{2}>0)\wedge\left(  {\Delta R(x)}<x_{1}\leq{\Delta S(x)}\right)  \right\}%
\end{align}
where $\mathbf{p=}\min\left\{{\Delta D(x)},{\Delta C(x)}\right\}$.

\item \textit{Closing in} domain: the speed difference is large and the distance is not. $Dom((h,w_{4}))$ is the set%
\begin{align}\label{Dom_q4}
&\left\{ x\in X:(x_{2}\leq0)\wedge({\Delta R(x)}<x_{1}\leq{\Delta S(x)})\right\}\\
\nonumber &\cup \left\{x\in X:(x_{2}=0)\wedge(x_{1}={\Delta R(x)})\right\}
\end{align}%

%
\item \textit{Danger} domain: the follower distance from the leader vehicle is close to the unsafe one. $Dom((h,w_{5}))$ is the set%
\begin{align}\label{Dom_q5}
\left\{ x\in X:\left(  {\Delta E(x)}\leq x_{1}\leq{\Delta R(x)}\right)  \right\}\backslash \left\{x\in X:(x_{2}=0)\wedge(x_{1}={\Delta R(x)})\right\}
\end{align}
%

\item \textit{Unsafe} domain: a collision with the leader will occur if the leader deceleration is maximum.
\begin{equation}\label{Dom_q6}%
Dom((h,w_{6}))=\left\{x\in X:x_{1}<{\Delta E(x)}\right\}
\end{equation}%

\end{enumerate}

\subsection{External events}\label{sub_sec_external_events}
The set of events $V$ is defined as
\begin{equation} \label{event_set}%
V=\left\{  \sigma_{nl},\sigma_{ex},\sigma_{c}\right\}
\end{equation}
The $\sigma_{nl}$ event indicates that a new leader appears in front of the current vehicle.
When the vehicle is in a no lane change mode, this can happen because the former leader left the current lane or because a vehicle from another lane starts to move to the current lane, in front of the current vehicle. 
\begin{assumption}\label{assumption_time_sigma_nl}
The interval of time between any two occurrences of the event $\sigma_{nl}$ is greater than a known finite nonnegative value.
\end{assumption}

The event $\sigma_{c}$ is associated to the crossing of the midline of a lane, or of the separation line between two lanes. It will be used in the situation where the following vehicle must consider a new leader (the one in the new lane) or stop the lane change maneuver, as explained in Section \ref{subsec_reset}.


Given $n$, let us consider $\mathcal{I}_n(t)=\{m,i,j\}$. Suppose that $m\neq \epsilon$. For a given $T>0$, let us define the utility functions
\begin{equation}\label{eq_utility_generic_r}
{\widehat{U}}(t) =\int_{t-T}^t  min(v_{des}^n,\widehat{v}(\tau))d\tau
\end{equation}
and
\begin{equation}\label{eq_utility_generic_l}
{\widetilde{U}}(t) =\int_{t-T}^t  min(v_{des}^n,\widetilde{v}(\tau))d\tau
\end{equation}
where $\widetilde{v}(\tau)=v^j(\tau)$ if $j\neq\epsilon$ and $\widetilde{v}(\tau)=v_{max}$ otherwise. The time $T$, introduced to have short term averaged measurements of the involved velocities, is measured in seconds. The utility functions (\ref{eq_utility_generic_r}) and (\ref{eq_utility_generic_l}) are associated respectively to the current lane and to the target lane. 

The symbol $\sigma_{ex}$ denotes the event corresponding to a lane change being the best option with respect to safety and utility functions (\ref{eq_utility_generic_r}), (\ref{eq_utility_generic_l}) considerations, as indicated to be needed in \cite{A_toledo_2003} and \cite{A_Helbing2007_MOBIL}.

Moreover, to allow lane change, the vehicle needs to take into account the distances from the vehicles $i$ and $j$. Such distances must be large enough for the safety condition to be satisfied, i.e. to brake safely.
%
If $i\neq\epsilon$ and $j\neq\epsilon$ let us consider the distances
\begin{equation}\label{Delta_Lf}%
{\Delta \widetilde{L}} =\left\{
\begin{array}
[c]{l}%
s+c_{r}\frac{v^{n}}{a_{\max}}{v}^j+\Delta p\\
s+\frac{1}{2 a_{\max}}\left(2c_{r}v^{n}{v}^j + ({v}^j-v^{n})^2\right) +\Delta p 
\end{array}
\right.
\begin{array}
[c]{c}%
{v}^j>v^{n}\\
{v}^j\leq v^{n}
\end{array}
\end{equation}
\begin{equation}\label{Delta_Lf2}%
{\Delta \overline{L}} =\left\{
\begin{array}
[c]{l}%
s+c_{r}\frac{v^{n}}{a_{\max}}{v}^i\\
s+\frac{1}{2 a_{\max}}\left(2c_{r}v^{n}{v}^i + ({v}^i-v^{n})^2\right)
\end{array}
\right.
\begin{array}
[c]{c}%
{v}^i>v^{n}\\
{v}^i\leq v^{n}
\end{array}
\end{equation}
where, for notational simplicity, time has been omitted.

The parameter $\Delta p$ is calculated at current time $t$, in such a way that, when $n$ switches from the current to the target lane, its distance from the vehicle $j$ is greater than the risky distance, by assuming that $j$ will start braking at time $t$ with maximum deceleration.

Finally, the last condition for the event to be generated is that the vehicle is not already in a lane change mode, i.e. $\phi=0$. It means that once the lane change process has started, it cannot be stopped: the
vehicle moves to the target lane even if the previous lane is becoming more
performing.
In summary, the event $\sigma_{ex}$ is generated at time $t$ whenever the following conditions hold: 
\begin{equation}\label{eq_utility_generic_cond_micro}
{\widehat{U}}(t)\leq {\widetilde{U}}(t) - c
\end{equation}
\begin{equation}\label{eq_sicurezza_avanti}
({p}^j-p^{n}\geq\Delta \widetilde{L})\wedge (j\neq \epsilon)
\end{equation}

\begin{equation}\label{eq_sicurezza_indietro}
(p^{n}-{p}^i\geq\Delta \overline{L})\wedge (i\neq \epsilon)
\end{equation}

\begin{equation}\label{eq_sicurezza_phi}
\phi=0
\end{equation}
where $c$ is a fixed cost associated to the lane change.

\subsection{Edges}
\begin{figure}[]
	\centering\includegraphics[width=0.8\columnwidth]{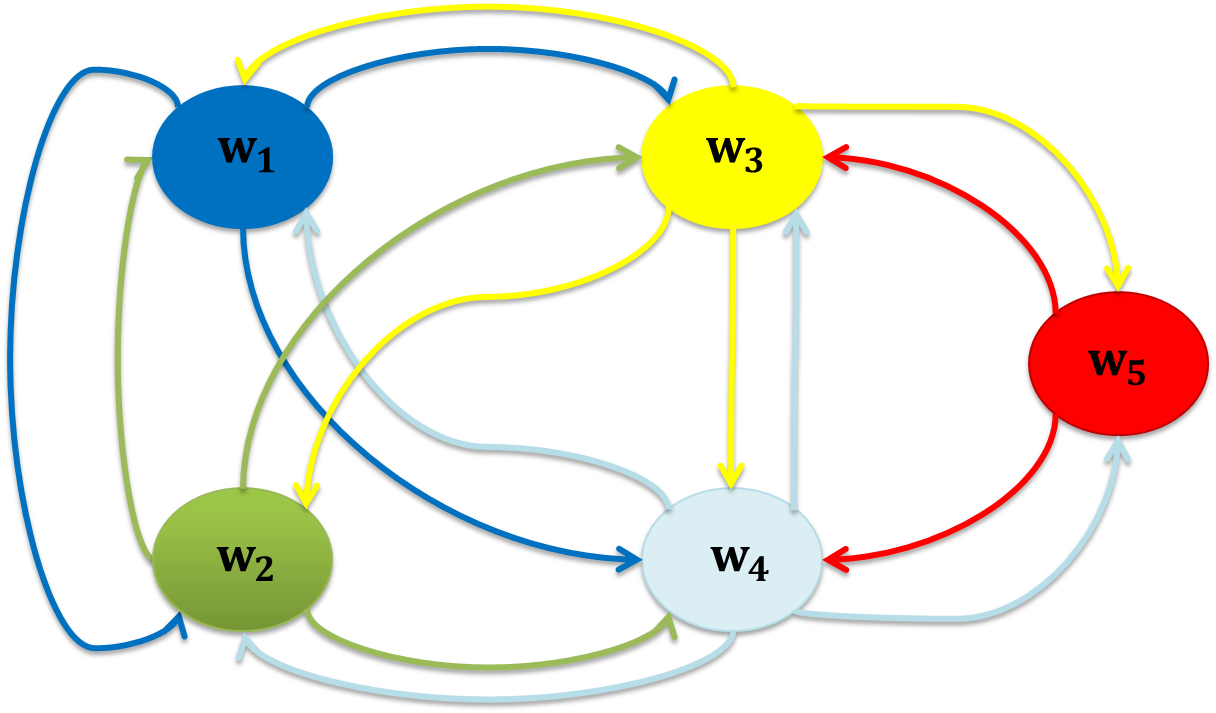}
	\caption{The discrete states considered in $\mathcal{E}"$ and their transitions: colors are related to the diagram depicted in Figure \ref{Figure_DeltaX_V_diagram}.}\label{Figure_discrete_transitions}
\end{figure}

Let us now describe the set of edges $\mathcal{E}$. Let $\mathcal{E}^{\prime
}\subset Q_{1}\times V\times Q_{1}$ denotes the set of edges in the cycle
(\ref{ciclo}).%
\begin{align}
\mathcal{E}^{\prime}=\{ &  (r,\sigma_{ex},r2l),(l,\sigma_{ex},l2r),(r2l,\sigma
_{c},l2l),\label{eq_set_cycle}\\
&  (l2l,\sigma_{c},l),(l2r,\sigma_{c},r2r),(r2r,\sigma_{c},r)\}\nonumber
\end{align}
The symbol $\mathcal{E}"$ denotes the "low level" set of state-dependent
transitions described in Figure \ref{Figure_discrete_transitions}. In this
case there is no external event causing the transition. Hence an element
of $\mathcal{E}"$ is denoted by $\left(  w_{i},\epsilon,w_{j}\right)$, with $\epsilon$ being the null event,
or simply $\left(  w_{i},w_{j}\right)  $. Let
\begin{align}
\mathcal{E}"=\{ &  (w_{1},w_{2}),(w_{1},w_{3}),(w_{1},w_{4}),(w_{2}%
,w_{1}),(w_{2},w_{3}),\\
&  (w_{2},w_{4}),(w_{3},w_{1}),(w_{3},w_{2}),(w_{3},w_{4}),(w_{3}%
,w_{5}),\nonumber\\
&  (w_{4},w_{1}),(w_{4},w_{2}),(w_{4},w_{3}),(w_{4},w_{5}),(w_{5}%
,w_{3}),\nonumber\\
&  (w_{5},w_{4})\}\nonumber
\end{align}
Note that the mode $w_{6}$ never appears since it represents a critical
situation and the control laws are defined in such a way that, by construction, no critical situation is reached.

Finally,%
\begin{equation}
\widehat{\mathcal{E}}=\left\{  \left(  w_{i},\sigma_{nl},w_{j}\right)
:i,j\in\left\{  1,...,5\right\}  \right\}  \label{eq_set_edges_sigma1}%
\end{equation}

At low level we assume that the transitions in $\widehat{\mathcal{E}}$ have
priority over the transitions in $\mathcal{E}"$.

The transitions of the overall system are denoted by $\mathcal{E}\subseteq Q\times  2^{V} \times Q$. A transition in $\mathcal{E}$ occurs if a transition occurs at high level or at low level. We assume that event triggered transitions always have
priority over state triggered ones. The concurrency of more than
one event in $V$ can be taken into account. In fact, let us consider the following conditions:
\begin{align}
a &  :\left(  h_{i},\sigma^{\prime},h_{j}\right)  \in\mathcal{E}^{\prime}\\
b &  :w_{f}=w_{g}\\
c &  :\left(  w_{f},w_{g}\right)  \in\mathcal{E}"\\
d &  :\left(  w_{f},\sigma_{nl},w_{g}\right)  \in\widehat{\mathcal{E}}\\
e &  :h_{i}=h_{j}%
\end{align}
Then
\begin{align}
\mathcal{E}=\{ &  \left(  \left(  h_{i},w_{f}\right)  ,\sigma,\left(
h_{j},w_{g}\right)  \right)  :\left(  a\wedge b\wedge\sigma=\left\{
\sigma^{\prime}\right\}  \neq\left\{  \sigma_{nl}\right\}  \right)  \vee\\
\nonumber &  \left(  c\wedge e\wedge\sigma=\epsilon\right)  \vee\left(  d\wedge
e\wedge\sigma=\left\{  \sigma_{nl}\right\}  \right)  \vee\\
\nonumber &  \left(  a\wedge d\wedge\sigma=\left\{  \sigma^{\prime},\sigma_{nl}\right\}
\right)  \vee\left(  a\wedge b,\forall\sigma^{\prime}\in\left\{  \sigma
_{ex},\sigma_{c}\right\}  \wedge\sigma=\left\{  \sigma_{ex},\sigma
_{c}\right\}  \right)  \vee\\
\nonumber &  \left(  a\wedge b,\forall\sigma^{\prime}\in\left\{  \sigma_{ex},\sigma
_{c}\right\}  \wedge d\wedge\sigma=\left\{  \sigma_{ex},\sigma_{c},\sigma
_{nl}\right\}  \right)  \}
\end{align}
\subsection{Continuous dynamics}
For $q=\left(h,w\right)\in Q$, the continuous evolution is determined by the function $f_{q}$, where 
\begin{align}
f_{q}(x,d) =&\left[
\begin{array}
[c]{cccccccc}%
0 & 1 & 0 & 0 & 0 & 0\\
0 & 0 & 0 & 0 & 0 & 0\\
0 & 0 & 0 & 0 & 0 & 0\\
0 & 0 & 0 & 0 & 1 & 0\\
0 & 0 & 0 & 0 & 0 & 0\\
0 & 0 & 0 & 0 & 0 & 0
\end{array}
\right]  {x}-\left[
\begin{array}
[c]{c}%
0\\
cos(x_6)\\
0\\
0\\
sin(x_6)\\
0
\end{array}
\right]  g_{q}({x})+\left[
\begin{array}
[c]{c}%
0\\
1\\
1\\
0\\
0\\
0
\end{array}
\right] d
\end{align}
%
Let $u(t)=g_{q}(x(t))$, with $\left\vert u(t)\right\vert \leq a_{\max}$, be the state feedback control input. 
\subsection{Control laws}
We now introduce the state feedback control laws that depend on the discrete state. In the following description $\alpha_{1}$, $\alpha_{2}$, $\alpha_{4}$, $\tau$, $b_d$ are positive sensitivity parameters, while $v_{des}$ is the desired speed the driver wants to achieve. 

\begin{enumerate}%
\item $q=(h,w_1)$, $h\in\{l,r\}$: the follower vehicle can run freely because the leader vehicle is either too far away or faster or both. The acceleration will depend only on actual and desired speed.
%
\footnotesize
\begin{align}\label{q1control_action}
g_{q}({x}(t))= \max\left\{  \alpha_{1}\left( v_{des}-\left(  x_{3}(t)-x_{2}(t)\right)\right)  ,\varepsilon\ast sign(v_{des}-\left(  x_{3}(t)-x_{2}(t)\right) \right\}%
\end{align}
\normalsize
The positive parameter $\varepsilon\in\left(0, \frac{a_{max}}{\lambda}\right)$ is introduced in order to achieve finite time convergence to the equilibrium points (see \cite{B_khalil_2002}).%

\item $q=(h,w_1)$, $h\in\{l2r,l2l,r2l,r2r\}$:
As for $h\in\{l,r\}$, the vehicle can run freely because the leader vehicle is either too far away or faster or both. Its acceleration is now given by the MOBIL model in \cite{A_Helbing2007_MOBIL}:
\begin{equation}\label{q7control_action}%
g_{q}({x}(t))=a_{max}\left[1 - \left(\frac{x_{3}(t)-x_{2}(t)}{v_{des}}\right)^4\right]
\end{equation}
%

%
\item $q=(h,w_2)$, $h\in\{l,r\}$:
Here the driver is closing in on the vehicle. The input depends on relative
speed and distance, following a modified version of the model in
\cite{Gazis1961FollowTheLeader}, i.e. 
\footnotesize
\begin{equation}\label{q2control_action}%
g_{q}({x}(t))=\alpha_{2}\cdot\frac{\left(  v_{des}+x_{2}(t)\right)
}{G-x_{1}(t)}\left(  x_{3}(t)-x_{2}(t)\right)
\end{equation}
\normalsize
Here $G$ is a distance s.t. $G-x_{1}(t)>0$, $\forall \: x \in Dom(q)$.%
\item $q=(h,w_2)$, $h\in\{l2r,l2l,r2l,r2r\}$:
As for the case with $h\in\{l,r\}$, here the driver is closing in on the vehicle and the input depends on relative
speed and distance; the MOBIL model \cite{A_Helbing2007_MOBIL} is used, i.e. 
\footnotesize
\begin{align}\label{q8control_action}%
g_{q}({x}(t))=a_{max}\left[1 - \left(\frac{x_{3}-x_{2}}{v_{des}}\right)^4 - \left(\frac{s_n + (x_{3}-x_{2})*\tau+ \frac{x_{3}-x_{2}}{2\sqrt{a_{max}b_d}}}{x_{1}} \right)^2\right]
\end{align}
\normalsize

\item $q=(h,w_3)$, $h\in\{l,r,l2r,l2l,r2l,r2r\}$: the follower does not take any action, either because its speed is close to the leader's one and the distance is small or because the speed difference is too large with respect to the distance:
\begin{equation}\label{q3control_action}%
g_{q}({x}(t))=0
\end{equation}

%
\item $q=(h,w_4)$, $h\in\{l,r\}$: the speed difference is large and the distance is not, so a deceleration action is needed; it will be done following a nonlinear GHR model depending on distance and relative speed \cite{Gazis1961FollowTheLeader}:
\footnotesize
\begin{align}\label{q4control_action}
g_{q}({x}(t))= \min\left\{  -\alpha_{4}\cdot\frac{x_{3}^{2}
(t)-(x_{3}(t)-x_{2}(t))^{2}}{2\left(  x_{1}(t)+s+\frac{c_{s}\lambda}%
{a_{max}}x^{2}_{3}(t)\right)  },\varepsilon\ast sign(x_{2})\right\}%
\end{align}
\normalsize

where the positive parameter $\varepsilon$ is as in (\ref{q1control_action}).

\item $q=(h,w_4)$, $h\in\{l2r,l2l,r2l,r2r\}$: as for the case where $h\in\{l,r\}$, a deceleration is needed. This will be done according to the MOBIL model in \cite{A_Helbing2007_MOBIL}:
\footnotesize
\begin{align}\label{q10control_action}%
&g_{q}({x}(t))= -a_{max} \left(\frac{s_n + (x_{3}-x_{2})*\tau+ \frac{x_{3}-x_{2}}{2\sqrt{a_{max}b_d}}}{x_{1}} \right)^2
\end{align}
\normalsize

\item $q=(h,w_5)$, $h\in\{l,r,l2r,l2l,r2l,r2r\}$: the distance from the leader vehicle is close to the unsafe one and the follower uses the maximum deceleration:
\begin{equation}
g_{q}({x}(t))=-a_{\max}\label{u_q5}%
\end{equation}\label{q5control_action}%
\end{enumerate}
\subsection{Continuous Hybrid State evolution. Reset conditions}\label{subsec_reset}
Let $[t',t")$ be the time interval between two consecutive event triggered transitions. By definition, for a given $h\in Q_1$, $Dom((h,w_f))\cap Dom((h,w_g))=\emptyset$, $\forall f\neq g$.

Therefore we can define the function $\boldsymbol{I}:X\rightarrow \left\{ 1,2,...,6\right\} $, where $\boldsymbol{I}(x)=i:x\in Dom((h_{i},w_f))$. Then, given $d:\mathbb{R}\rightarrow \mathbb{R}$ and $x(t')$, the evolution in time of $\mathcal{H}$ is described by the pair
of functions $x:\mathbb{R}\rightarrow X$, $q:\mathbb{R}\rightarrow Q$, where
$x(t)$ is the solution of the equation
\begin{equation}
\dot{x}(t)=f_{\boldsymbol{I}(x(t))}(x(t),d(t)),t\in [t',t")
\end{equation}
with initial state $x(t')$ and
\begin{equation}
q(t)=q_{\boldsymbol{I}(x(t))},t\in [t',t")
\end{equation}
Therefore, to define the time evolution of the hybrid automaton, only reset conditions for event triggered transitions have to be defined. The symbol $x^{-}$ denotes the value of the state before the transition, i.e. $x^{-}=\lim_{t\rightarrow t"^-}x(t)$, while $x^{+}$ the state after the transition, i.e. $x(t") = x^+$ and $x(0) = x_0$ for a given initial continuous state $x_0$.
%
%

Let us consider a transition
\begin{align}\label{eq_transition_for_guard}
e = ((h_i,w_f), \sigma,(h_j,w_g))\in \mathcal{E}
\end{align}

According to the different events, different situations occur:
\begin{description}
\item[i)] $\sigma=\{\sigma_{nl}\}$.
$h_i=h_j$. Given the distance with the new leader $\Delta_{L}$ and its speed $v_{L}$%
\begin{equation}\label{eq_reset_nl}
x^{+}=\mathcal{R}\left(  e,x^{-},\left(  \Delta_{L},v_{L}\right)  \right)
\end{equation}
where
\begin{align}
x_{1}^{+}  & =\Delta_{L}\\
x_{2}^{+}  & =x_{3}^{-}-x_{2}^{-}+v_{L}\\
x_{3}^{+}  & =v_{L}\\
x_{i}^{+}  & =x_{i}^{-},i=4,5,6
\end{align}

\item[ii)] $\sigma=\{\sigma_{ex}\}$.
$\left(  h_{i},h_{j}\right)  \in\{  \left(l,l2r\right)  ,\left( r,r2l\right)\}$.
If $h_{i}=l$ and $h_j= l2r$, then
\begin{equation}\label{eq_reset_Llr}
x^{+}=R_{l}x^{-}
\end{equation}
\begin{equation}\label{eq_reset_Llr_Rl}
\mathcal{R}_l=\left(
\begin{array}
[c]{cccccc}%
1 & 0 & 0 & 0 & 0 & 0\\
0 & \cos\phi & 1-\cos\phi & 0 & 0 & 0\\
0 & 0 & 1 & 0 & 0 & 0\\
0 & 0 & 0 & 1 & 0 & 0\\
0 & \sin\phi & -\sin\phi & 0 & 0 & 0\\
0 & 0 & 0 & 0 & 0 & -\phi
\end{array}
\right)
\end{equation}
If $h_i=r$ and $h_j=r2l$, then
\begin{equation}\label{eq_reset_RrL}
x^{+}={R}_{r}x^{-}
\end{equation}
where $R_{r}$ is obtained from $R_{l}$ in (\ref{eq_reset_Llr_Rl}) by changing $\phi$ with $-\phi$.%
\item[iii)] $\sigma=\{\sigma_{c}\}$. Two cases are possible:
\item[a)] $\left(  h_{i},h_{j}\right)  \in\{  \left(r2l,l2l\right)  ,\left(l2r,r2r\right)\}$;
The external event $\sigma_{c}$ is generated by the crossing of the separation line between lanes $r$ and $l$, or $l$ and $r$ respectively. This transition implies a change of the leader for the vehicle $n$. Given the distance with the new leader $\Delta_{L}$ and its speed $v_{L}$, the state $x^{+}$ is defined as in (\ref{eq_reset_nl}). If no leader is defined, then  $v_{L}=v_{des}^n$ and $\Delta_{L}=\Delta_{max}$.
  \item[b)] $\left(  h_{i},h_{j}\right)  \in\{  \left(l2l,l\right)  ,\left(  r2r,r\right)\}$;
The external event $\sigma_{c}$ is generated by the reaching of the midline of lane $l$ or $r$, respectively.
%
%
\begin{equation}\label{eq_reset_llLrrR1}
x_{i}^{+} = x_{i}^{-},i\in\left\{  1,3,4\right\}
\end{equation}
\begin{equation}\label{eq_reset_llLrrR2}
x_{2}^{+}=x_{3}^{-}-\sqrt{\left(  x_{3}^{-}-x_{2}^{-}\right)  ^{2}+\left(  x_{5}^{-}\right)^{2}}
\end{equation}
\begin{equation}\label{eq_reset_llLrrR5}
x_{5}^{+}=0
\end{equation}
\begin{equation}\label{eq_reset_llLrrR6}
x_{6}^{+}=0
\end{equation}
The state after the transition takes into account the conservation of linear momentum.
\item[iv)]
In case of multiple events occurring at the same time, a hierarchy is needed for implementing all the conditions. When $\sigma_{c}$ occurs at the same time as another event, it has priority; then, for the other two events, $\sigma_{ex}$ has priority over $\sigma_{nl}$.
\begin{enumerate}
  \item[a)] $\sigma=\{\sigma_{ex},\sigma_{nl}\}$; first the conditions regarding $\sigma=\{\sigma_{ex}\}$ are implemented, then the ones regarding $\sigma=\{\sigma_{nl}\}$.
  \item[b)] $\sigma=\{\sigma_{c},\sigma_{nl}\}$; the conditions regarding $\sigma=\{\sigma_{c}\}$ are implemented before the ones regarding $\sigma=\{\sigma_{nl}\}$.
  \item[c)] $\sigma=\{\sigma_{c},\sigma_{ex}\}$; the conditions regarding $\sigma=\{\sigma_{c}\}$ are implemented before the ones regarding $\sigma=\{\sigma_{ex}\}$.
  \item[d)] $\sigma=\{\sigma_{c},\sigma_{ex},\sigma_{nl}\}$; first the conditions regarding $\sigma=\{\sigma_{c}\}$ are implemented, then the ones regarding $\sigma=\{\sigma_{ex}\}$ and finally the ones regarding $\sigma=\{\sigma_{nl}\}$.
\end{enumerate}
\end{description}
\subsection{$Init$ set}\label{subsec_init_set}
Finally, let us define the $Init$ set.%
\footnotesize
\begin{equation}\label{Init_qxDom}%
Init=\left( \bigcup_{h\in Q_1} \bigcup_{i=1}^{5}\{(h,w_i)\}\times\{Dom((h,w_i))\}\right)
\end{equation}
\normalsize

Note that the unsafe domain is not included in the $Init$ set. 
\subsection{Properties}\label{sec_properties}
The first property is related to the automaton $\mathcal{H}^{n}$. Considering every possible acceleration of the leader vehicle, 
\begin{proposition}\label{p1}
the hybrid automaton ${\mathcal{H}^{n}}$ is non-blocking, deterministic and non Zeno $\forall \: d:\mathbb{R}\rightarrow\mathbb{R}$.
\end{proposition}%
\begin{pf}
Consider the time interval $[t',t")$ between two consecutive event triggered transitions. Then, by construction (see the definition of domains and control laws) each hybrid state evolution in this time interval is non-blocking and deterministic. By analyzing the direction of the vector $\dot{x}$ along the boundaries of the domains, we can state that any evolution in $[t',t")$  is non Zeno.
Moreover, because of Assumption \ref{assumption_time_sigma_nl}, and of the physical meaning of the events $\sigma_{c}$ and $\sigma_{nl}$, in each finite time interval a finite number of event triggered transitions can occur. Therefore the hybrid automaton is non-blocking, deterministic and non Zeno.
\end{pf}

Let us now consider the cluster of vehicles.
\begin{definition}
Given a decentralized control law, the cluster of controlled vehicles $\{1,...N\}$ is safe if for any initial condition in $(Init)^{N}$ collisions are avoided $\forall \: t\geq t_0$, with respect to the worst case scenario.
\end{definition}
\begin{proposition}\label{Prop_8}
The cluster of controlled vehicles $\{1,...N\}$ described by $\mathcal{H}^{n}$, $n\in\{1,...N\}$, is safe.
\end{proposition}
\begin{pf}
When in lane maintain mode, a collision between $n$ and its leader is avoided by design. Therefore, by definition of leader, collision is avoided with any other vehicle. When $n$ is in lane changing mode, a collision with any vehicle in its neighborhood is avoided by definition of the $\sigma_{ex}$ generation mechanism. Therefore collision is avoided with any other vehicle.
\end{pf}

The cluster described by $\mathcal{H}^{n}$, $n\in\{1,...N\}$, is an autonomous system, where all the control laws have been defined. Hence Proposition \ref{Prop_8} states that safety is assured if all the controllers apply the appropriate control law. 
Therefore safety is assured in the worst case scenario, as required in Problem definition.

Let us now analyze the stability properties of the cluster. The case of a single lane has been investigated in \cite{Iovine2015ADHSofficial} and corresponds to the assumption that no event occurs.

Let $x^{n}(t)\in X$ be the continuous state of $\mathcal{H}^{n}$. Let $X_{e}^{n}\subset X$ be the set
%
\begin{align}\label{eq_equilibrium_set_n}
X_{e}^{n}=S_1^n \cup S_2
\end{align}
where
\begin{align}\label{eq_equilibrium_set_n_S1}
S_1^n= \{ {x}\in X: (x_3-x_2=v_{des}^{n}) \wedge (x_5=0) \wedge (x_6=0)\}
\end{align}%
\begin{align}\label{eq_equilibrium_set_n_S2}
S_2= \{ {x}\in X: (x_{2}=0) \wedge (\Delta R(x)\leq x_{1}\leq\Delta S(x)) \wedge(x_5=0) \wedge (x_6=0)\}
\end{align}%
Define $\mathbf{x}(t)$ and $\mathbf{X}_{e}$ as:
\begin{equation}\label{eq_x_del_platoon}
\mathbf{x}(t)=x^{1}(t)\times x^{2}(t)\cdots\times x^{N}(t)\in X^{N}%
\end{equation}
and
\begin{equation}\label{eq_x_eq_del_platoon}
\mathbf{X}_{e}=X_{e}^{1}\times X_{e}^{2}\cdots\times X_{e}^{N}\subset X^{N}%
\end{equation}
\begin{proposition}\label{p3}
There exists $\widehat{t}$ such that, for any initial condition in $(Init)^{N}$, 
 $\mathbf{x}(t)\in\mathbf{X}_{e}$, $\forall t\geq\widehat{t}$.
\end{proposition}

The statement in Proposition \ref{p3} is due to finite time convergence of the control laws in (\ref{q1control_action}) and (\ref{q4control_action}).
Moreover, if $v_{des}^i=v_{des}^j$ $\forall \: i,j=1,...,N$ or $v_{des}^i\neq v_{des}^j$ $\forall \: i\neq j$ then $\mathbf{x}(t)\in S_1^1\times S_1^2 \times ... \times S_1^N$. Therefore the second requirement in Problem \ref{problem} definition is met in these cases. In general, vehicle $n$ may reach $v_{des}^n$ depending on the initial conditions in the cluster.

Let us note that the conditions that generate the event $\sigma_{ex}$ do not allow lane changing when the closest vehicles ahead on the two lanes have exactly the same speed: we dot not address this pathological case here.
%

%
%
%
\section{Mesoscopic model}\label{Mesoscopic model}
In real life, the human behaviour is related to macroscopic quantities, such as traffic density. Being variance a density-dependent function (see \cite{Helbing1999}), a variance-driven adaptation mechanism is adopted in order to improve the overall system performance.

%
Inspired by \cite{Helbing2006}, in \cite{Iovine2015ADHSofficial} a variance-driven time headways (VDT) mechanism is applied to the hybrid automaton, in the single lane case. In this paper the same variance-driven mechanism applies also to utility functions, which determine lane changes. Considering the time headways  $T_R(x)$, $T_S(x)$ (see (\ref{headways})) and $T_D$ (see (\ref{Delta_D})), let us define new time headways by scaling the old ones with a suitable time varying parameter $\alpha_{T}(t)$, i.e.
\begin{align}\label{newhead}
&T'_R(x,t)={\alpha_{T}}(t)T_R(x), \:\: T'_S(x,t)={\alpha_{T}}(t)T_S(x), \:\:\\
\nonumber &T'_E(x,t)=T_E(x), \:\: T'_D={\alpha_{T}}(t)T_D
\end{align}%
where $T'_E(x,t)$ is not affected by the multiplication factor because of its meaning. The parameter ${\alpha_{T}}(t)$ is a global measurement of the current state of the traffic flow. More specifically,
$$
\alpha_{T}(t)\in\lbrack \alpha_{T}^{0},\alpha_{T}^{\max}]
$$
where $\alpha_{T}^{\max}$ is determined
from empirical data of the time-headway distribution for free and congested
traffic (see \cite{Helbing2006}). The value $\alpha_{T}^{0}$  is a sort of "safety margin" (if $\alpha_{T}(t)=0$ were possible, then the control law of vehicle $n$ would be "free driving" until the emergency distance from the leader would be reached. Then, in the worst case of maximum deceleration of the leader, the follower $n$ would brake with maximum intensity, which is not a desired behaviour). The value of $\alpha_{T}(t)$ is given by:
\begin{equation}\label{alfaT}
{\alpha_{T}}(t)=sat(1+{z}(t))
\end{equation}%

where, denoting with $H_L$, $card(H_L)=h_L$,  the set of vehicles in front of $n$, in the same lane of $n$, within a given maximum distance,
\begin{equation}\label{zetaFORalfa}%
z(t)=\int_{t-T}^{t}\gamma V_{n}(\tau) sign\left[  v^n(\tau) - \left(\frac{1}{h_L}\sum_{i\in H_L}{v^{i}}(\tau)\right)\right]
d\tau
\end{equation}
 The parameter $\gamma$ is a sensitivity parameter determined from empirical data and $V_{n}$ is the "variation coefficient", function of mean speed and variance of the vehicles in $H_L$:
\begin{equation}\label{variation_coefficient}
V_{n}(t) = \frac{\sqrt[2]{\frac{1}{h_L}\sum_{i\in H_L}\left(  {v^{i}}(t)-\left(  \frac{1}{h_L}\sum_{i\in H_L}{v^{i}}(t)\right)
\right)^{2}}}{\frac{1}{h}\sum_{i\in H_L}{v^{i}}(t)}
\end{equation}
The parameter defined in (\ref{alfaT}) allows to take into account both increasing (case ${\alpha_{T}}(t)>1$) and decreasing velocity variations (case ${\alpha_{T}}(t)<1$).

Once ${\alpha_{T}}(t)$ is computed, new time varying domains are obtained by replacing the old time headways with the new ones.
Safety is still ensured, even in the worst case. Propositions \ref{Prop_8} and \ref{p3} still hold for the modified model.

By using the parameter ${\alpha_{T}}(t)$, the $\mathcal{H}^{n}$ model becomes able to take into account the neighboring environment: an anticipatory action will be suggested to the driver in the ADAS case, or will be provided autonomously in the ACC case.

In order to calculate the $\alpha_T$ parameter, on-line calculation of mean speed and variance is needed: this information is propagated by a vehicular network. Each vehicle will send data concerning its position and speed through such network. Those data have to be received in "real-time" (compared to the human response time and safety-critical time-response) in order to select the right control action. In Section \ref{vehicular_network} a feasibility study of the needed communication network is described, which will allow us to consider such hypothesis as feasible.
The introduced methodology is also applied to the definition of a mesoscopic level utility function. Let us define a parameter ${\tilde{\alpha}_{T}}(t)$ which is the same as ${\alpha_{T}}(t)$ but computed for the target lane. Then at mesoscopic level the condition (\ref{eq_utility_generic_cond_micro}) becomes:
\begin{equation}\label{eq_utility_meso}
\left({\widehat{U}}(t)\leq {\widetilde{U}}(t) - c\right) \vee \left[
\left({\widetilde{U}}(t) - c< {\widehat{U}}(t)\leq {\widetilde{U}}(t)\right) \wedge \left({\alpha_{T}}(t){\widehat{U}}(t)\leq {\tilde{\alpha}_{T}}(t){\widetilde{U}}(t)\right)\right]
\end{equation}

The utilization of the macroscopic level information helps in dealing with the pathological case of having the two ahead vehicles at the same speed; still, when all the ahead vehicles share the same speed the problem is still unsolved.

\section{Vehicular networks}\label{vehicular_network}
Nowadays connected-vehicles (see \cite{Uhlemann}), namely vehicles that include
interactive advanced driver-assistance systems (ADASs) and cooperative
intelligent transport systems (C-ITSs), are a reality and in the next years
will be part of our daily life. Just consider that the European Parliament
stated that by 31 March 2018 all new models of passenger cars and light
commercial vehicles will have to be equipped with the eCall system (i.e., the
European automatic emergency service).
Much interest is arising on C-ITSs, where vehicles cooperate by exchanging
messages wirelessly to achieve a higher level of safety and to avoid
on-the-road hazards. The cellular network (e.g., LTE)
is not a suitable choice
for safety applications due to stringent requirements for both bounded delay
and high reliability; consequently dedicated protocol stacks have been developed.
In this section we give a brief look at vehicular technologies and standards that can allow the operation of the proposed solution.
\subsection{Vehicular technologies}\label{vehicular_tecno}
WAVE (Wireless Access in Vehicular Environment) is the protocol stack defined
by the IEEE with the intent of extending the 802.11 family to include
vehicular environments. The early standards were approved in 2010 and,
commonly, WAVE refers to the IEEE 802.11p and IEEE 1609.x standards.

For the purpose of our application, it is important to remark the following
characteristics of IEEE 802.11p:
\begin{itemize}
  \item operation range up to 1000 m;
  \item communications in high-speed and high-mobility scenarios;
  \item priority and power control.
\end{itemize}
Among all supported architectures, the Independent BSS (IBSS) network topology
allows a set of stations to directly communicate with each other. This
capability, also known as ad hoc networking, achieves connectivity everywhere
since it does not require a fixed infrastructure to establish the connections.

Safety applications do not require all the features defined in the TCP/IP
network and transport layers that would also introduce unwanted overhead and
delay; to support them, the WAVE Short-Message Protocol (WSMP) was defined and
standardized through IEEE 1609.3. WSMP messages are allowed in the WAVE
Control Channel and therefore can benefit from a higher transmission power and
a favorable scheduling time.
\subsection{The Challenge of Network Delay}\label{network_delay_problem}
Delay is a crucial parameter in real-time networks and services. The ITU-T
G.114 recommendation states that with a unidirectional end-to-end delay lower
than 150 ms, most applications, both speech and non-speech, will experience
essentially transparent interactivity.
For road safety applications, it is conventionally assumed that the maximum
end-to-end delay must be below 100 ms, while for traffic efficiency the
threshold rises to 500 ms.
On the other hand, vehicle drivers have different reaction times, depending on
specific stimuli. In \cite{Triggs1982} most unalerted drivers have shown
themselves capable of responding in less than 2.5 s in urgent situations.
Similar results can be derived from \cite{Ihata2002reactiontime}.
The reaction time is lower if the stimuli is not only visual, but it is also vibratory or auditory
(see \cite{Pratichizzo2013reactiontimes}), so an ADAS can inform the driver of an imminent
hazard, to increase the probability of an in-time response.

Simulations conducted in \cite{Mahonen2010} confirm that the Control Channel
has good characteristics in terms of End-to-End Delay and Radio Range for
safety messages (lower than 20 ms and higher than 750 m, respectively) when
the message rate is lower than 1000 packets/s and the message size is below
256 Bytes. Estimating the delay in a multi-hop scenario is not trivial, since
many variables (e.g., distance, interference, computing time) impact on it,
but with some assumptions we can respect an upper-bound value of 100 ms. The
proposed controller considers a vehicle as a predecessor or a follower only if
its distance is up to 500 m but the WAVE technology allows to send data more
than 750 m away (in \cite{Mahonen2010} the maximum distance reached is around
2.5 km in open air scenarios). Consequently the number of relays needed to
reach the last vehicle is not strictly bounded to the number of vehicles in
the queue but depends essentially on the distance. Considering clusters up to
5 vehicles, the worst case occurs when vehicles are along a line 500 m apart
from each other and the channel quality allows a maximum range of 750 m: to
reach the end of the line under such conditions, 4 relays are needed with a
total delay lower than 80 ms plus the eventual computation time on the relay
nodes. As the environment is highly dynamic, the topology changes over time
but, if the vehicles are still clustered, the delay and the number of hops are
surely better than in the worst case. It follows that the reduction of the
number of vehicles in a single cluster and the minimization of the processing
operations on the information to be relayed are sufficient to comply the
constraints of safety applications.

It is worth noting that, when more vehicles are within the Radio Range, the
same message may arrive several times to the same vehicle. Moreover, the
message arrival does not implicitly provide information about the positions of
vehicles that either generated or relayed it. The first problem is easily
solved including Timestamp and Serial Number in the packet in order to enable
the Duplicate Discovery Process; the second issue can be simply managed using
GPS geolocation. Each vehicle can include its own positioning data at the time
of generating a message (and leave them unchanged in the messages it relays)
and can build the physical topology collecting the localization information of the received messages.

\subsection{Lane Identification}\label{network_lane_id}
In order to support the multiple lanes scenario, the connected vehicles should also transmit the currently used lane and the total number of road lanes in the actual position. The former should be an integer ranging from 1 to $n$, where 1 is the rightmost lane and $n$ is the last lane on the left; obviously $n$ cannot be lower than 1. Along a route the total number of lanes can vary often so, since the current lane identifier is relative to the right boundary of the road, their total number should also be announced. With those two values is possible to satisfy a necessary condition to form a cluster: \textit{``only if two vehicles advertise the same values for the number of available lanes and the current one, then they are on the same lane and the ACC can be applied''}. For the sake of completeness, the previous condition can be strengthened with a statement about the estimated distance (through the GPS positions): \textit{`` if the same value of the current lane is advertised by a vehicle farther than another one that announces a different number of total lanes, then the road is not homogeneous and thus that information must be discarded''}.

In this extended context the problem of identifying the lane currently traveled by the vehicle must be solved. The use of the \textit{Global Positioning System}, guarantees measurements with an accuracy between 10 and 20 meters that isn't enough for the purpose of our study. The main sources of errors, that can be compensated only partially, affecting the GPS accuracy are described in the following.

\textbf{\textit{Dilution of Precision (DOP)}} -- The geometry of the satellite constellation can be such that the satellite sightlines can be close to collinear, providing a sub-optimal position solution; the only way to compensate this effect is to choose, when it's possible, one satellite near the perpendicular and the  other three equally distributed near the horizon.

\textbf{\textit{Multipath}} -- Nearby objects produce reflected and/or diffracted signals beside the direct line of sight component. In single frequency GPS receivers, multipath is a dominant error source in pseudorange measurements, as well as in differential carrier phase measurement (DGPS). The errors due to the multipath effect are estimated in \cite{KosELMAR2010} to 8 meters in average.

\textbf{\textit{Atmospheric affects}} -- A phase shift in the signal from the GPS satellites occurs when the signal passes through the charged particles of the ionosphere and the water vapor in the troposphere. The former effect is inversely proportional to the square of the frequency so, with the use of two frequencies (L1 -- 1575.42 $MHz$ and L2 -- 1227.60 $MHz$), it is possible to estimate and delete the ionospheric delay. The latter depends from the local humidity values and can be avoided only through a Satellite Based Augmentation System (SBAS) (e.g., \textit{WAAS} in the USA, \textit{EGNOS} in Europe, \textit{MSAS} in Japan); the key idea is to measure the humidity with ground stations and deliver this information with some additional geostationary satellites.

A receiver enabled to collect SBAS measures can estimate the position with an accuracy lower than 7.6m (the field measures in the USA, have proven the \textit{WAAS} capable of an accuracy between 1 and 6 meters, \cite{WideAreaAugmentation2015}); this improved accuracy is due to the atmospheric effects reduction and to the effective transformation of the system in a DGPS.
Since the SBAS satellites are near the horizon line, many ground terminals aren't able to receive their signal so, optionally, the same information can be provided through the Internet (e.g., the \textit{ESA SISNeT Project}, \cite{toran2001internet}). However, this solution does not provide the benefit of the differential system and therefore is less accurate than a DGPS.

In urban environments, where tall buildings, overpasses, and other obstructions to the sky are present, GPS and DGPS suffer from inaccuracies and outages that make them not appropriate for a vehicle exact localization. We exclude the use of active or passive elements along the roads (e.g., for magnetic sensing and radio frequency identification) since the proposed system is expected to work almost everywhere, hence a coverage of all the roads through sensors is inconceivable. The road lane identification can be facilitated by vision based systems and geospatial databases containing the road geometry and lane boundary positions (\cite{SkogITS2009}, \cite{McCallITS2006}, \cite{OgawaIVS2006}).

The mainly studied approach inside vision systems, relies on the processing of images obtained by an on-board camera to identify road elements. It's a cheap and versatile solution for the road lanes identification even if it suffers the variation of visibility conditions (e.g., illumination, weather, shadows, obstacles) and of the status of the road (e.g., absence of lane markings, painting quality).
The Light Detection and Ranging (LIDAR) technology is more reliable and uses ultraviolet, visible, or near infrared light to illuminate objects and then, from the analysis of the reflected light, build a 3D map of the surrounding environment. Compared to a camera sensor, the LIDAR is more expensive but its diffusion is increasing in the automotive industry and the prices are going down.
In conjunction with these technologies, it is possible to use maps and geospatial data to cross check the measures or to provide additional information on speed limits, intersections, and so on.

Devices as the ones introduced are responsible to generate the set of events $V$.
\section{Simulation results}\label{Simulation results}
In this section, we will show some simulation results for the developed hybrid model. Our aim is to verify the ability to correctly represent a safe car-following situation (collision avoidance) both in lane maintaining and in lane changing conditions and to compare the use or disuse of the VDT-like mechanism in a complex traffic scenario.

Each vehicle is supposed to be driven by an ACC system controlled by the proposed mesoscopic hybrid automaton.
\subsection{Single lane case}
In order to validate results, we first choose a scenario with a single lane road and a cluster composed by five vehicles, whose parameters are described in Table \ref{table_values} and initial conditions in Table \ref{table_initial_values_simulation}. Here a sort by location index can be used.

\begin{figure}
	\centering\includegraphics[width=0.8\columnwidth]{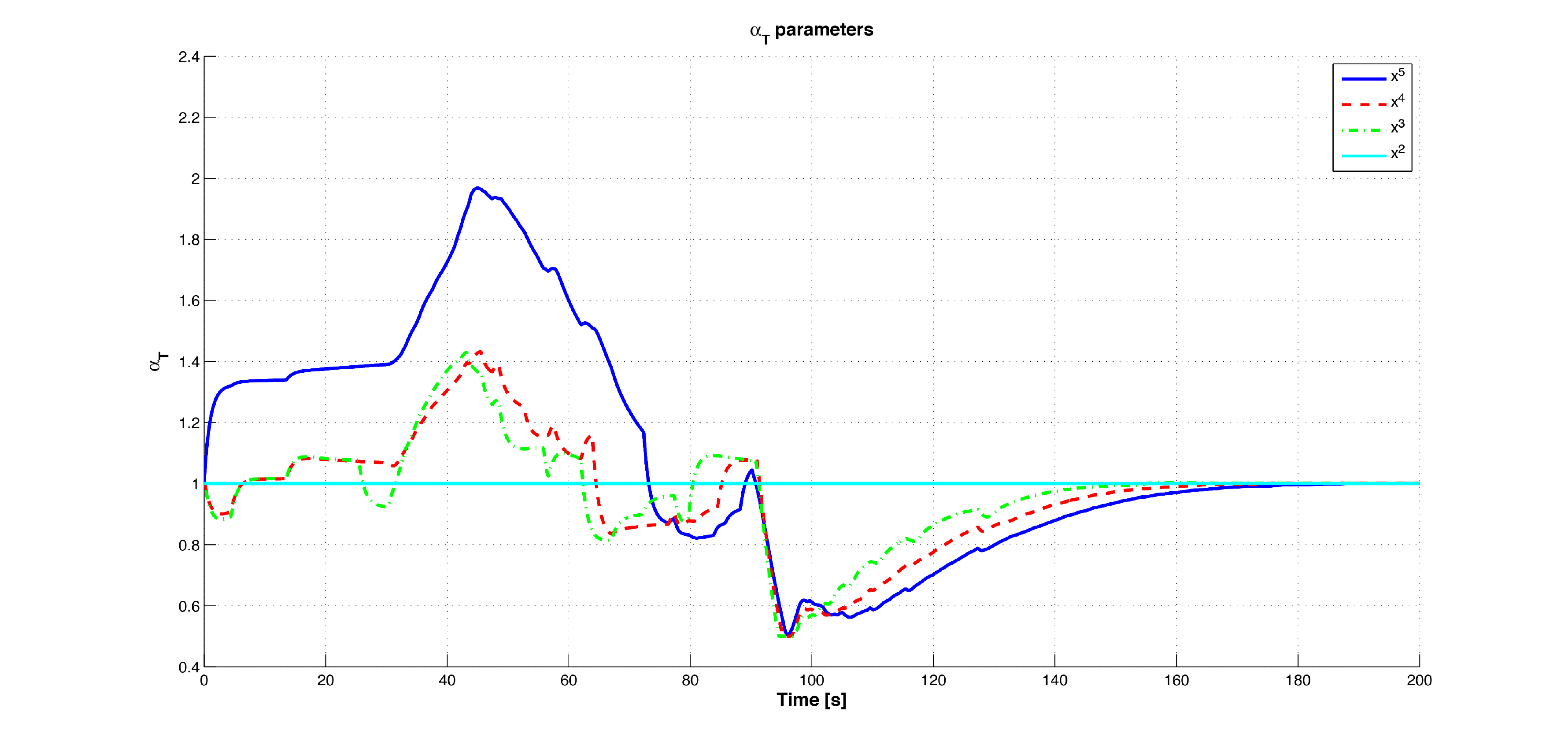}
  	\caption{The different values that the parameter $\alpha_T$ takes for each vehicle. The $\alpha_T$ corresponding to the fifth vehicle is described by the blue line.}\label{Figure_alphat}
\end{figure}

At the beginning, the first vehicle is the leader of a group composed by four vehicles: at time 30" it decelerates for reaching the velocity of 18 m/s, while at time 90" it speeds up to a velocity of 33 m/s. The fifth vehicle reaches and tag along to the group: its initial separation is the maximum we consider for an interaction between two vehicles (segment road of 500 meters), according to Section \ref{vehicular_network}. The technological constraints are then fulfilled. The fifth vehicle is approaching a group that is decreasing its initial speed: the $\alpha_T$ parameter utilization is expected to anticipate its deceleration. Both the active or inoperative VDT-like mechanism mode are implemented and compared; in the figures, the first situation is indicated with (b) case, while the second one with the (a) case. 
\begin{table}
  \centering
\begin{tabular}{|c|c|c|c|}
\hline \textbf{Parameter} & \textbf{Value} & \textbf{Parameter} & \textbf{Value}\\
\hline $L_{n}$ [m] & 4.5 & $L_{0}$ [m] & 0.5\\
\hline $\lambda$ & 2 & $a_{max}$ [m/$s^2$]& 5\\
\hline $c_{r}$ & 0.2 & $c_{s}$ & 0.2\\
\hline $c_{c}$ & 10 & $T_{D}$ [s] & 20\\
\hline $v_{max}$ [m/s] & 36 & $\alpha_{1}$ & 0.1\\
\hline $\alpha_{2}$ & 0.1 & $\alpha_{4}$ & 1\\
\hline $G$ [m] & 500 & $\varepsilon$ & 0.1\\
\hline $\Delta_{max}$ [m] & 500 & $\alpha^{max}_{T}$ & 2.2\\
\hline $\alpha^{0}_{T}$ & 0.2 & $\gamma$ & 4\\
\hline
\end{tabular}
\vspace{0.1cm}
  \caption{Parameters}\label{table_values}
\end{table}
\begin{table}
  \centering
\begin{tabular}{|c|c|c|c|}
\hline \textbf{Vehicle n} & \textbf{ $x_{1}(0)$} & \textbf{$x_{2}(0)$} & \textbf{$x_{3}(0)$}\\
\hline $x^{1}$ & 500 & 0 & 30\\
\hline $x^{2}$ & 50 & 0 & 30\\
\hline $x^{3}$& 50 & 0 & 30\\
\hline $x^{4}$ & 50 & 0 & 30\\
\hline $x^{5}$ & 500 & -6 & 30\\
\hline
\end{tabular}
\vspace{0.1cm}
  \caption{Initial values for single lane case}\label{table_initial_values_simulation}
\end{table}
Figure \ref{Figure_alphat} describes the continuous evolution of the $\alpha_{T}$ parameter: it raises when the group is decelerating and increases when is accelerating, as expected. The second vehicle takes into account only the first vehicle of the group; hence $\alpha_T$ will always be equal to 1. The case where $\alpha_T$ is not used can be represented by a value of $\alpha_T$ equal to 1 for all vehicles.

\begin{figure}
	\centering\includegraphics[width=0.8\columnwidth]{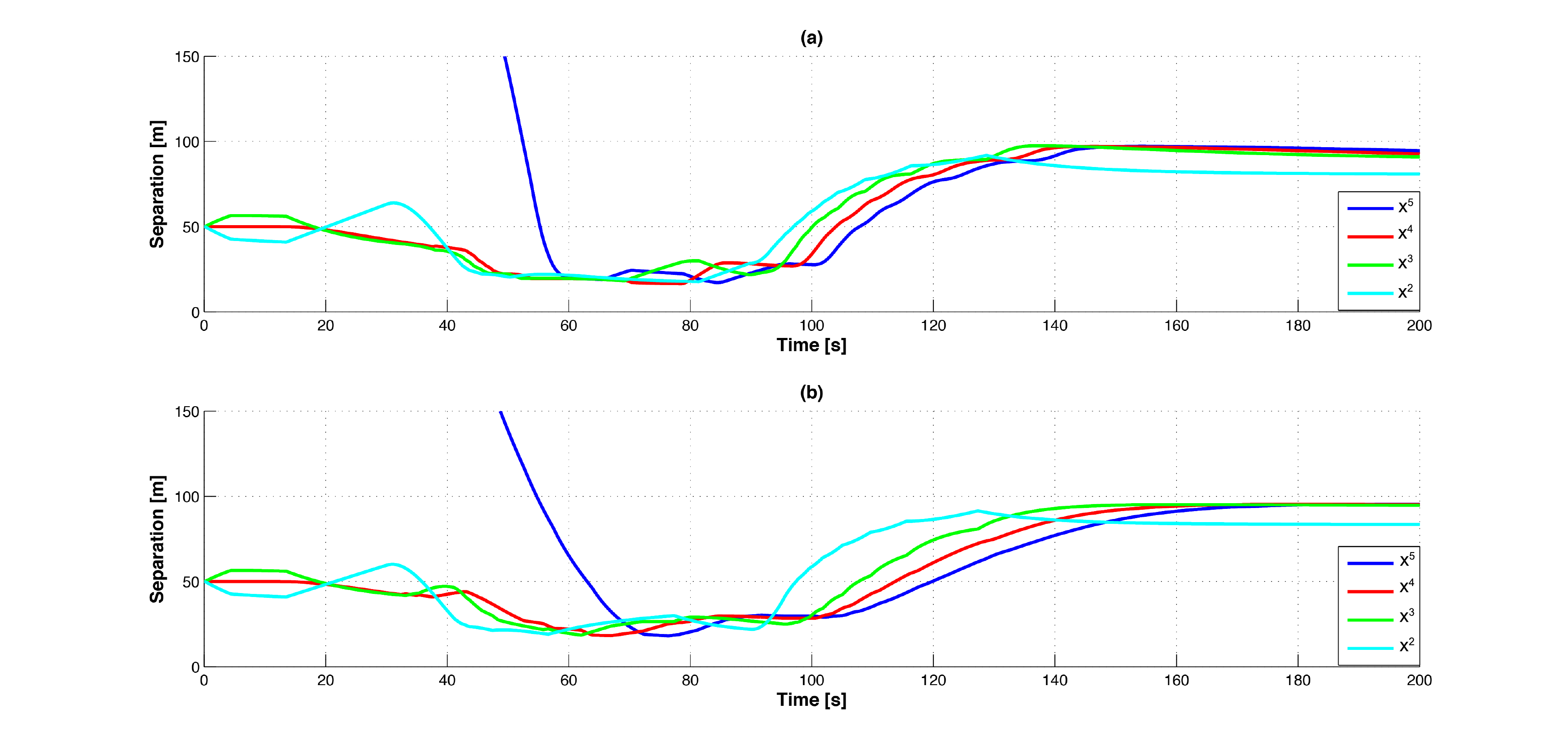}
	\caption{The separations among vehicles in the active VDT-like mechanism case, (b), and the inoperative one, (a). The fifth vehicle dynamics is depicted in blue color.}\label{Figure_separation_simulation}
\end{figure}

\begin{figure}
	\centering\includegraphics[width=0.8\columnwidth]{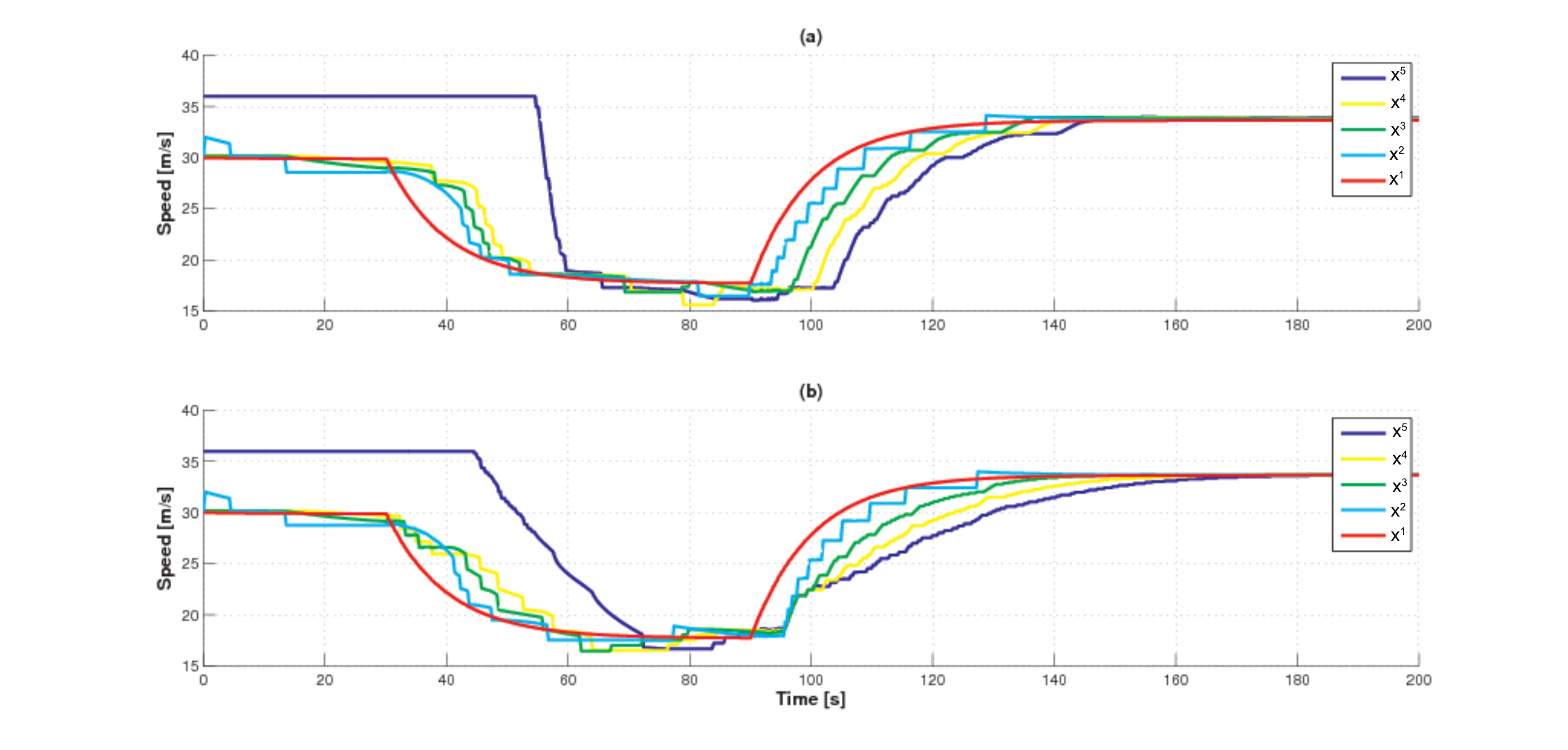}
    \caption{The speed of each vehicle in the active VDT-like mechanism case, (b), and the inoperative one, (a). The first vehicle dynamics is depicted in red, while the fifth one is in blue.}\label{Fig_comparison_sino_alphat}
\end{figure}

\begin{figure}
	\centering\includegraphics[width=1\columnwidth]{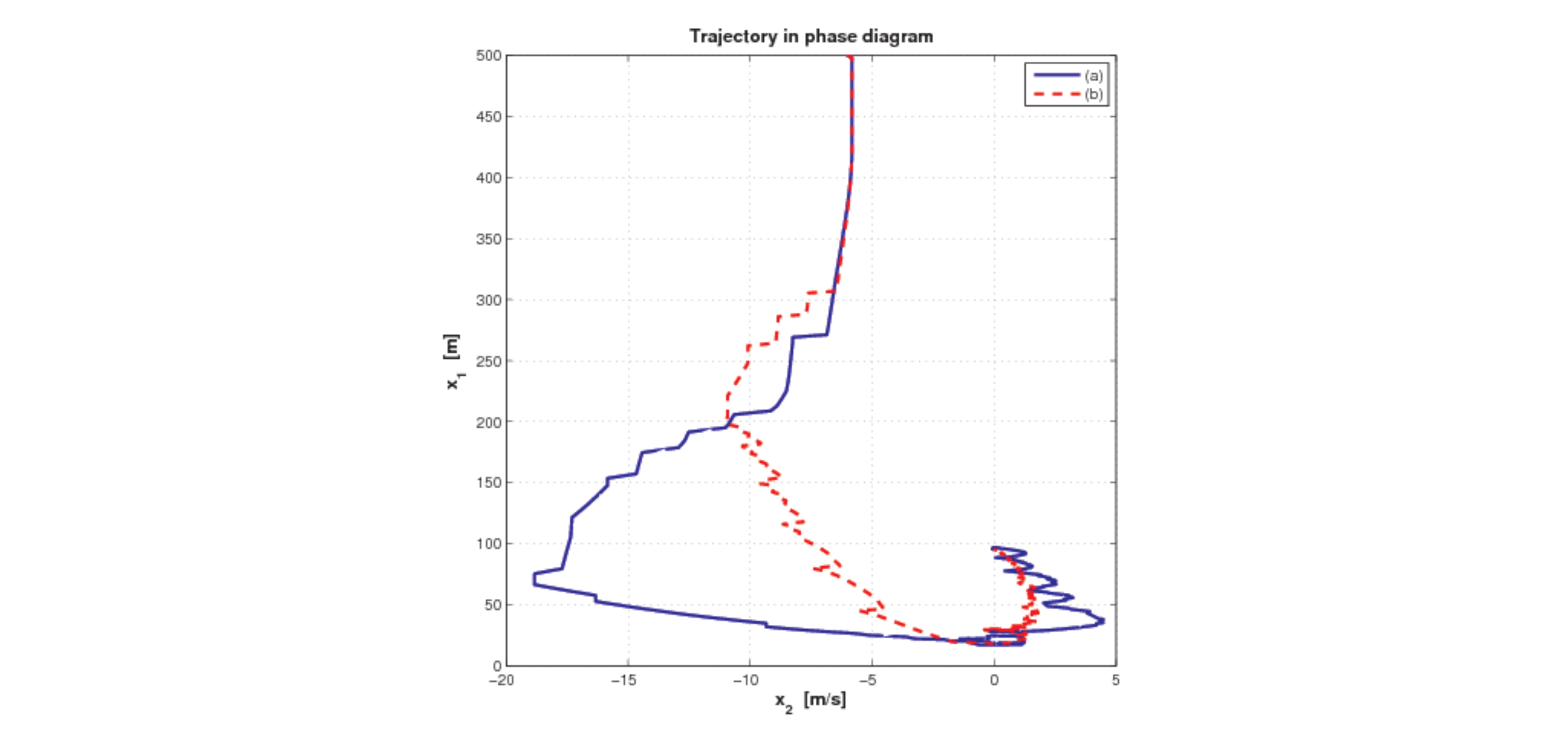}
	\caption{Trajectories in phase portrait are depicted: in (a) (blue curve) the group of vehicles does not use VDT-like mechanism, while in (b) (dotted curve) it does.}\label{Figure_phase_dynamics}
\end{figure}

Figure \ref{Figure_separation_simulation} shows the separations among vehicles: there is no collision in both cases. Also an anticipatory action is introduced when the VDT mechanism  is used. In
Figure \ref{Fig_comparison_sino_alphat} the anticipatory action of the VDT is clearly shown: in (b) the fifth vehicle starts decelerating before than in (a), around 45" and 55", respectively, as it can be seen from the shape of the blue curves.
A similar anticipatory action is seen when acceleration occurs: in (b), the fifth vehicle accelerates before the point representing 100", while in (a) it does after that point.
This leads to a smoother behavior, which is described in the $\mathcal{H}^{5}$ model phase portrait in Figure \ref{Figure_phase_dynamics}. %
%

It is possible to see that some oscillations take place before all vehicles reach their equilibria, as expected. In the simulations, less oscillations appear in the case where VDT-like mechanism is used. 
These results are important because non-smooth transients are responsible for string instability and shock waves propagation. Hence, less oscillations and less magnitude of oscillations, as shown in our results, provide a better behavior w.r.t these problems. 
In Figure \ref{Figure_phase_dynamics} steady-state behaviours are similar, while the transient is smoother in the case of VDT-like mechanism utilization.

%

\subsection{Two lanes case}
To validate the lane change case, we choose a scenario composed by five vehicles on each lane on a road with two lanes. The leaders of the groups share horizontal position and velocity, respectively on the $r$ and $l$ lanes. A sixth vehicle will follow the group on the $r$ lane, $x^{11}$. Table \ref{table_initial_values_simulation_lane_change} introduces the initial conditions for this scenario: for all vehicles, $x_5(0)=x_6(0)=0$. 

\begin{figure}
	\centering\includegraphics[width=0.8\columnwidth]{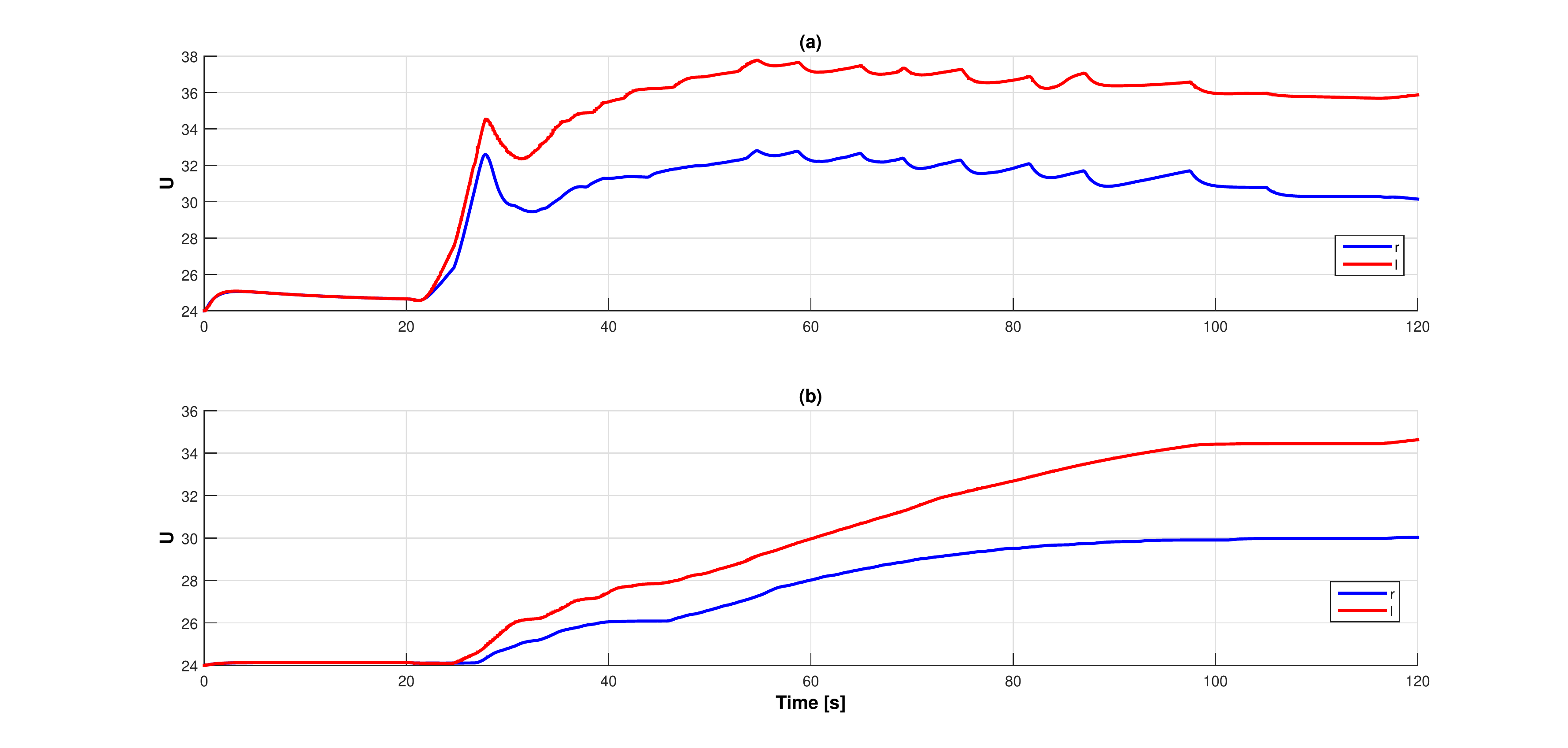}
    \caption{The utility functions $\widehat{U}$ and $\widetilde{U}$ for $r$ and $l$ lane, respectively the blue and red curves, according to mesoscopic description, (a), and microscopic one, (b).}\label{Fig_comparison_utility_lc}
\end{figure}

\begin{table}
  \centering
\begin{tabular}{|c|c|c|c|c|}
\hline \textbf{Vehicle} & \textbf{ $x_{1}(0)$} & \textbf{$x_{2}(0)$} & \textbf{$x_{3}(0)$}& \textbf{$x_{4}(0)$}\\
\hline $x^{1}$ & 500 & 0 & 24 & 2\\
\hline $x^{2}$ & 45 & 0 & 24 & 2\\
\hline $x^{3}$& 45 & 0 & 24 & 2\\
\hline $x^{4}$ & 45 & 0 & 24 & 2\\
\hline $x^{5}$ & 45 & 0 & 24 & 2\\
\hline $x^{11}$ & 45 & 0 & 24 & 2\\
\hline $x^{6}$ & 500 & 0 & 24 & 7\\
\hline $x^{7}$ & 45 & 0 & 24 & 7\\
\hline $x^{8}$ & 45 & 0 & 24 & 7\\
\hline $x^{9}$ & 45 & 0 & 24 & 7\\
\hline $x^{10}$ & 45 & 0 & 24 & 7\\
\hline
\end{tabular}
\vspace{0.1cm}
  \caption{Initial values for lane change case}\label{table_initial_values_simulation_lane_change}
\end{table}
At the beginning the group leaders have the same velocity; then at time 20" both accelerate up to reach a new velocity according to the different limitations imposed on the lanes. The leader of the $l$ lane will have a higher velocity target than the leader of the $r$ one; then all vehicles on the $l$ lane can gain from this situation and reach a higher velocity than the one of the vehicles on the $l$ lane. The event $\sigma_{ex}$ represents this gain and allow the lane change because safety is ensured.
Figure \ref{Fig_comparison_utility_lc} describes the behaviour of the utility functions. 

\begin{figure}
	\centering\includegraphics[width=0.8\columnwidth]{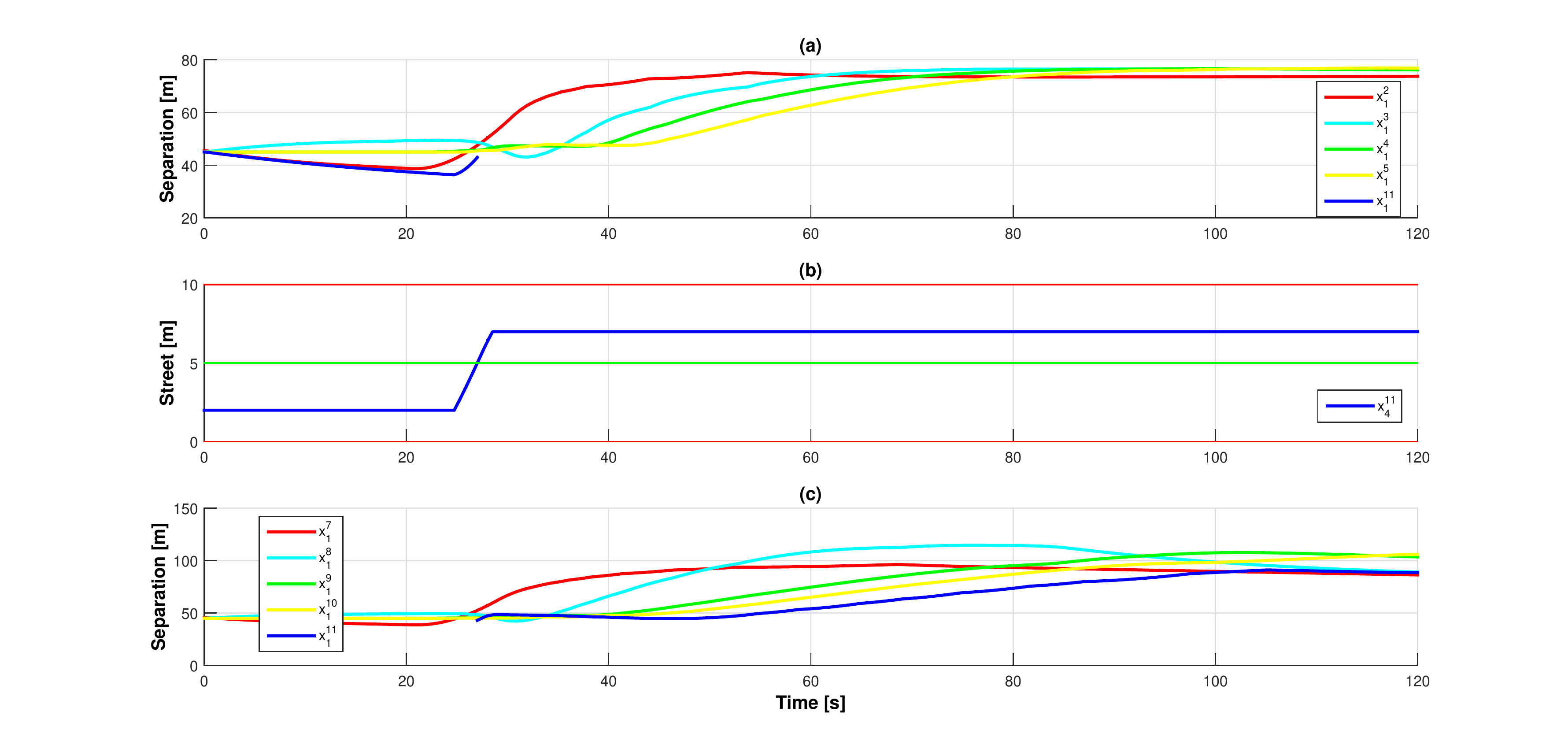}
    \caption{The separations among vehicles in the two lanes; (a) represents the $r$ lane and (c) represents the $l$ lane, while (b) describes the motion on the $\vec{y}$ axis for $x^{11}$.}\label{Fig_comparison_distances_lc}
\end{figure}

When the conditions generating $\sigma_{ex}$ are matched, the $x^{11}$ vehicle starts a lane change and overtakes the group of vehicles on the $r$ lane, as shown in Figure \ref{Fig_comparison_distances_lc}. The blue curves represent the dynamics of $x^{11}$ on both the longitudinal axis and $\vec{y}$, and on both the lanes. Notice that the blue curve in (a) stops before the end of the simulation and that the one in (c) starts after the beginning of the simulation; this is due to the lane change.
Figure \ref{Fig_comparison_distances_lc} confirms that no collisions occur and that the equilibrium distances among vehicles are reached.

\begin{figure}
	\centering\includegraphics[width=0.8\columnwidth]{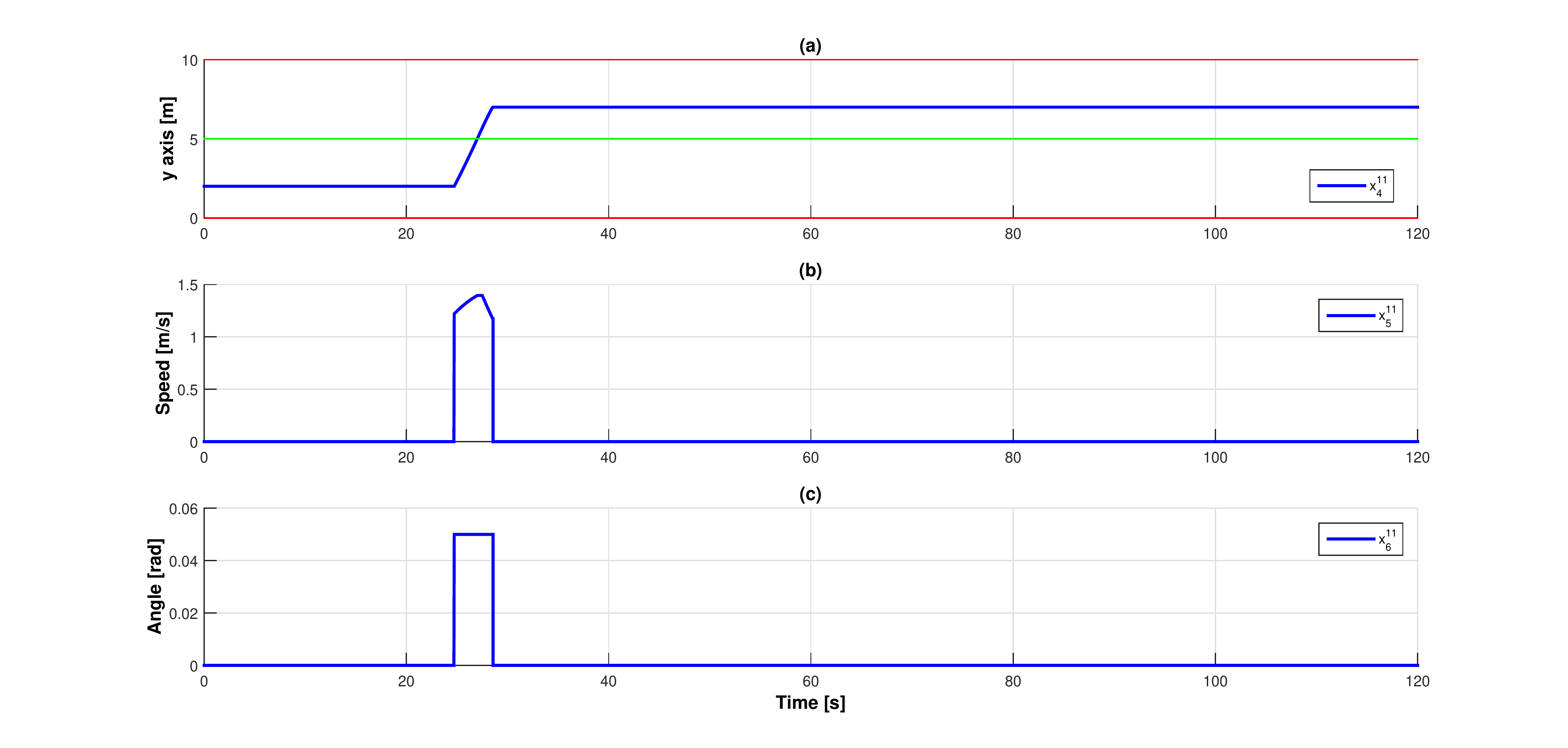}
    \caption{The position and velocity of $x^{11}$ on the $\vec{y}$ axis and the needed angle.}\label{Fig_comparison_x4x5x6_lc}
\end{figure}

Only during the lane change, there is a dynamics on the position and velocity of $x^{11}$ on the $\vec{y}$ axis; furthermore, the value of the angle is different from zero (see Figure \ref{Fig_comparison_x4x5x6_lc}).

\begin{figure}
	\centering\includegraphics[width=0.8\columnwidth]{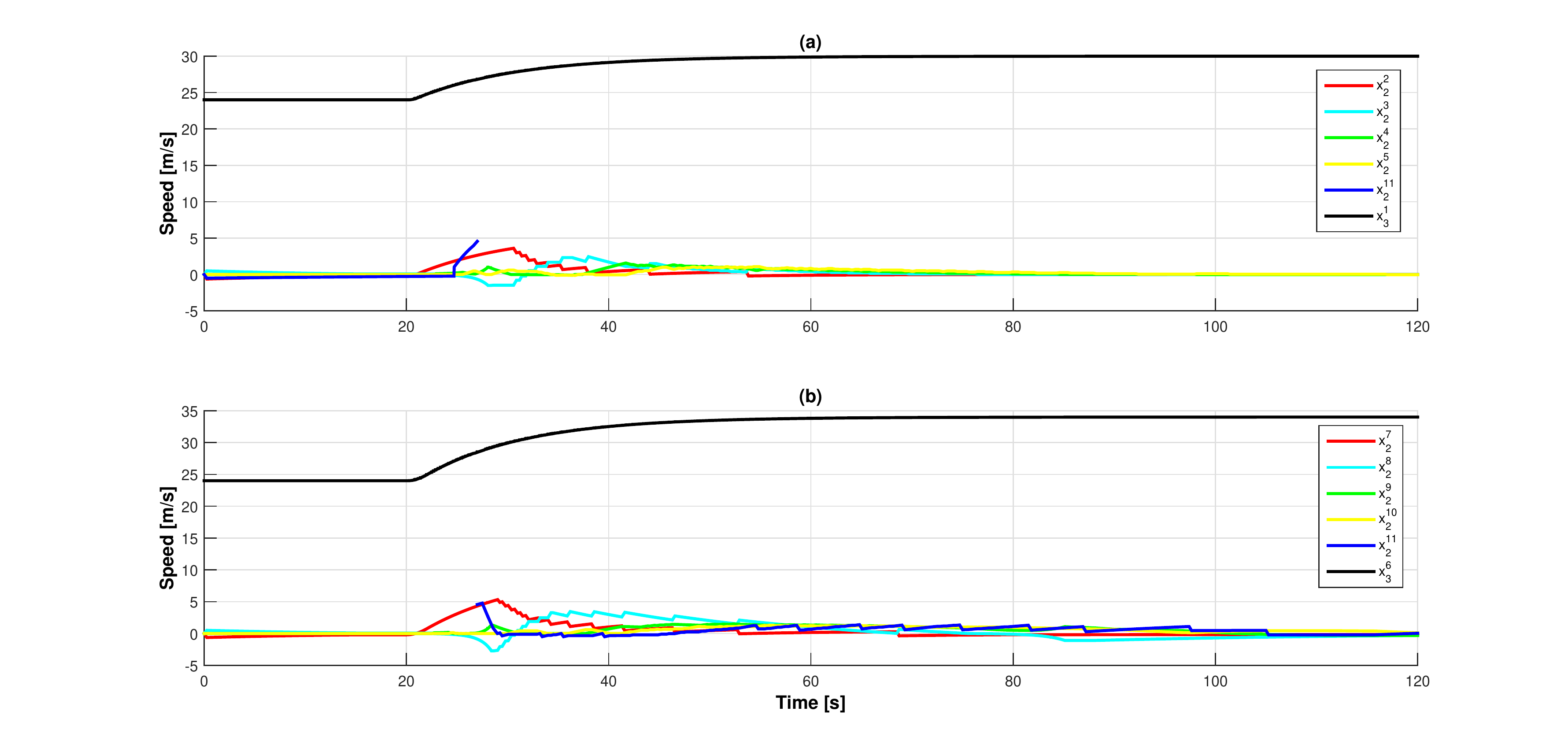}
    \caption{The differences of velocity between each pair of vehicles in the $r$ lane, (a), and in the $l$ lane, (b). The blue curves refer to the $x^{11}$ vehicle, while the black ones are the velocities of the leaders.}\label{Fig_comparison_speeds_lc}
\end{figure}

Figure \ref{Fig_comparison_speeds_lc} introduces the differences of velocity between each pair of vehicles in the $r$ lane, (a), and in the $l$ lane, (b). It also shows the velocity of the leaders of the two lanes.

The simulations show how the introduced ACC model fits the desired target of no collision event, both in lane maintaining and in lane change situations. Furthermore, they also illustrate how ACC performances are improved by taking into account information about the neighbouring vehicles.
\section{Conclusions}\label{Conclusions}
A novel mesoscopic hybrid automaton for car-following situations is introduced: its purpose is to safely control the single vehicle dynamics through an Adaptive Cruise Control that replaces human control actions. The controller processes other vehicles information and takes decisions about braking or throttle actions on the basis of a human-inspired model. 
The controller does not literally follow human behavior since weaknesses, such as mistakes or distractions, are not included. However, adaptability to various conditions and comfort are featured. 
%

The hybrid automaton has been proven to satisfy safety and stability properties.
To our best knowledge, no hybrid automaton has been presented in the literature as a description of a microscopic human-driven vehicle. Furthermore, to better represent  human behavior, a macroscopic value dependence is added to the final model: the consideration of macroscopic quantities is an important step-ahead to be considered, both in the single lane and multi-lane situations.
Feasibility of the needed communication network is analyzed and described. Finally simulation results confirm traffic throughput improvements using the proposed scheme.

%
%
\nocite{ArtHybrid}
\nocite{B_slotine_li_1991}
\bibliographystyle{abbrv}
\bibliography{biblio_}
%
\end{document}